\documentclass{tlp}

\usepackage{float,epsfig}
\usepackage{epsf}
\usepackage{subfigure}
\usepackage{latexsym}
\usepackage{fancybox}

\newcommand{\mycomment}[1]{}

\newcommand{\cA}{{\cal A}}
\newcommand{\cB}{{\cal B}}
\newcommand{\cC}{{\cal C}}
\newcommand{\cF}{{\cal F}}
\newcommand{\cH}{{\cal H}}
\newcommand{\cI}{{\cal I}}
\newcommand{\cJ}{{\cal J}}
\newcommand{\cL}{{\cal L}}
\newcommand{\cN}{{\cal N}}
\newcommand{\cO}{{\cal O}}

\newcommand{\cS}{{\cal S}}

\newcommand{\bottom}{\perp}

\newcommand{\mif}{\mbox{ :- }}

\newcommand{\wfsmeth}{{\sc Abdual}}
\newcommand{\unfr}{{\sc co-unfounded set removal}}

\newcommand{\longline}{\noindent\rule{\textwidth}{.01in}}

\newcommand{\newsg}{{\sc New Subgoal}}
\newcommand{\pgmcr}{{\sc Program Clause Resolution}}
\newcommand{\anscr}{{\sc Answer Clause Resolution}}

\newcommand{\delay}{{\sc Delaying}}
\newcommand{\simpl}{{\sc Simplification}}

\newtheorem{theorem}{Theorem}[section]

\newtheorem{lemma}[theorem]{Lemma}

\newtheorem{example}{Example}[section]
\newtheorem{dummylemma}{Dummy}[section]

\newtheorem{ExampleN}{\bf Example}[section]

\newenvironment{definition}
               {\begin{DefinitioN}~\rm}{\hspace{1cm} \hfill $\Box$\end{DefinitioN}}

\newtheorem{JDefinitioN}[section]{\bf Definition}

  {\begin{list}{} %
	{\setlength{\leftmargin}{10pt}%
         \setlength{\rightmargin}{10pt}%
	}%
  }%
  {\end{list}}

\floatstyle{ruled}
\newfloat{Algorithm}{thpb}{lop}[section]

\newtheorem{DefinitioN}[dummylemma]{\bf Definition}%[section]

\newcommand{\cE}{{\cal E}}
\newenvironment{CProg}{\begin{center} \tt \begin{tabular}[c]{l}}{\end{tabular} \end{center}}

\begin{document}

\title[Abduction in Well-Founded Semantics and Generalized Stable Models] 
        {Abduction in Well-Founded Semantics and Generalized Stable Models 
        via Tabled Dual Programs}

\author[Jos\'e J\'ulio Alferes, Lu\'\i{}s Moniz Pereira and Terrance Swift]
{Jos\'e J\'ulio Alferes \\
        A.I. Centre, Faculdade de Ci\^{e}ncias e Tecnologia \\
        Univ. Nova de Lisboa, 2825-516 Caparica, Portugal \\
        \email {jja@di.fct.unl.pt} 
\and 
Lu\'\i{}s Moniz Pereira \\ 
        A.I. Centre, Faculdade de Ci\^{e}ncias e Tecnologia \\
        Univ. Nova de Lisboa 2825-516 Caparica, Portugal \\
        \email{lmp@di.fct.unl.pt} 
\and
Terrance Swift \\
        Department of Computer Science \\
        SUNY Stony Brook, Stony Brook, NY. USA \\
        \email{tswift@cs.sunysb.edu}
}

\sloppy
\maketitle

\begin{abstract}
Abductive logic programming offers a formalism to declaratively
express and solve problems in areas such as diagnosis, planning,
belief revision and hypothetical reasoning.  Tabled logic programming
offers a computational mechanism that provides a level of
declarativity superior to that of Prolog, and which has supported
successful applications in fields such as parsing, program analysis,
and model checking.  In this paper we show how to use tabled logic
programming to evaluate queries to abductive frameworks with integrity
constraints when these frameworks contain both default and explicit
negation.  
% TLS minor wording change.
The result is the ability to compute abduction over well-founded
semantics with explicit negation and answer sets.  Our approach
consists of a transformation and an evaluation method.  The
transformation adjoins to each objective literal $O$ in a program, an
objective literal $not(O)$ along with rules that ensure that $not(O)$
will be true if and only if $O$ is false.  We call the resulting
program a {\em dual} program.  The evaluation method, \wfsmeth, then
operates on the dual program.  \wfsmeth{} is sound and complete for
evaluating queries to abductive frameworks whose entailment method is
based on either the well-founded semantics with explicit negation, or
on answer sets.  Further, \wfsmeth{} is asymptotically as efficient as
any known method for either class of problems.  In addition, when
abduction is not desired, \wfsmeth{} operating on a dual program
provides a novel tabling method for evaluating queries to ground
extended programs whose complexity and termination properties are
similar to those of the best tabling methods for the well-founded
semantics.  A publicly available meta-interpreter has been developed
for \wfsmeth{} using the XSB system.
\end{abstract}

%%%%
Submitted: May 3, 2001, revised: April 24, 2002, Dec. 1, 2002,
accepted: May 7 2003.
%%%%

\section{Introduction}

Abductive logic programming (see e.g. \cite{KaKT93}) is a general
non-monotonic formalism whose potential for applications is striking.
As is well known, problems in domains such as diagnosis, planning, and
temporal reasoning can be naturally modeled through abduction.  In
this paper (which is an extended and revised version with proofs of
\cite{AlPS99}), we lay the basis for efficiently computing queries over
ground three-valued abductive frameworks that are based on extended
logic programs with integrity constraints, and whose notion of
entailment rests on the well-founded semantics and its partial stable
models. Both the generalized stable models semantics \cite{KaMa90} and
the answer set semantics \cite{GeLi90} are also captured, their
two-valuedness being imposed by means of appropriate integrity
constraints.

Our query processing technique, termed \wfsmeth{}, relies on a mixture
of program transformation and tabled evaluation. In our abductive
framework, a transformation removes default negative literals from
both the program over which abduction is to be performed and from the
integrity rules.  Specifically a {\em dual transformation} is used,
that defines for each objective literal $O$ and its set of rules $R$,
a dual set of rules whose conclusion $not(O)$ is true if and only if
$O$ is false by $R$.  Tabled evaluation of the resulting program turns
out to be much simpler than for the original program, whenever
abduction over negation is needed.  At the same time, termination and
complexity properties of tabled evaluation of extended programs are
preserved by the transformation when abduction is not needed.

Regarding tabled evaluation, \wfsmeth{} is in the line of SLG
evaluation \cite{CheW96} which computes queries to normal programs
according to the well-founded semantics. In fact, its definition is
inspired by a simplification, for ground programs, of SLG as
reformulated in \cite{Swif99b}. To it, \wfsmeth{} tabled evaluation
adds mechanisms to handle abduction\footnote{Namely, by adding
abductive contexts to goals, by modifying operations on forests to
deal with such contexts, and by having a new operation to abduce
literals.}, and to deal with the dual programs\footnote{Namely by
introducing a \unfr{} operation.}.

The contributions of this paper are:
\begin{itemize}
\item We describe \wfsmeth{} fully and first consider its use over
abductive frameworks whose entailment method is based on the
well-founded semantics with explicit negation.  \wfsmeth{} is sound,
complete, and terminating for queries to such frameworks over finite
ground programs and integrity rules.  Furthermore, \wfsmeth{} is
ideally sound and complete for countably infinite ground programs.

%As a special case of this
%result, \wfsmeth{} is sound, complete, and terminating for finite
%ground and extended programs.
%
\item We show that over abductive frameworks whose entailment method
is based on the well-founded semantics with explicit negation, the
complexity of \wfsmeth{} is in line with the best known methods.  In
addition, for normal and extended programs --- viewed as abductive
frameworks containing no abducibles or integrity constraints --- query
evaluation has polynomial data complexity.
\item We provide a transformation that allows \wfsmeth{} to compute 
generalized stable models and answer sets under a credulous semantics,
and show that \wfsmeth{} provides a sound and complete evaluation
method for computing such models.  Furthermore, the efficiency of
\wfsmeth{} in computing generalized stable models is in line with the
best known methods.
\item Finally, we provide access to an \wfsmeth{} meta-interpreter,
written using the XSB system, illustrating how to evaluate \wfsmeth{}
in practice and describe how \wfsmeth{} can be applied to medical
diagnosis \cite{GSTPD00}, to reasoning about actions \cite{ALPPP00},
and to model-based diagnosis of an electric power grid \cite{CasP02}.
\end{itemize}

\section{Preliminaries}\label{sec:prelim}

\subsection{Terminology and assumptions} \label{sec:terminology}
Throughout this paper, we use the terminology of Logic Programming as
defined in, e.g.  \cite{Lloy84}, with the following modifications.  An
{\em objective literal} is either an atom $A$, or the {\em explicit
negation} of $A$, denoted $-A$.  If an objective literal $O$ is an
atom $A$, the explicit conjugate of $O$ ($conj_E(O)$) is the atom
$-A$; otherwise if $O$ has the form $-A$, the explicit conjugate of
$O$ is $A$.  A {\em literal} either has the form $O$, where $O$ is an
objective literal, or $not(O)$ the {\em default negation} of $O$.  In
the first form, where a literal is simply an objective literal, it
is called a {\em positive literal}; in the second, where it is of the
form $not(O)$, it is called a {\em negative literal}.  Default
conjugates are defined similarly to explicit conjugates: the default
conjugate ($conj_D(O)$) of an objective literal $O$ is $not(O)$, and
the default conjugate of $not(O)$ is $O$.  Thus, every atom is an
objective literal and every objective literal is a literal.
%Literals of the form $not(O)$ are called {\em default literals}.  
A {\em program} $P$ (sometimes also called an extended program),
formed over some countable language of function and predicate symbols
$\cL_P$, is a countable set of rules of the form $H \mif Body$ in
which $H$ is an objective literal, and $Body$ is a possibly empty
finite sequence of literals.  If no objective literals in a program
$P$ contain the explicit negation symbol, $P$ is called {\em normal}.
In either case, the closure of the set of literals occurring in $P$
under explicit and default conjugation is termed $literals(P)$, while
the closure of the set of objective literals occurring in $P$ closed
under explicit conjugation is termed $objective\_literals(P)$.

By a {\em three-valued interpretation\/} $\cI$ of a ground program $P$
we mean a subset of $literals(P)$.
% old \cH_P
We denote as $\cI_T$ the
set of objective literals in $\cI$, and as $\cI_F$ the set of literals
of the form $not(O)$ in $\cI$.  For a ground objective literal, $O$,
if neither $O$ nor $not(O)$ is in $\cI$, the truth value of $O$ is
undefined.  An interpretation $\cI$ is {\em consistent\/} if there is
no objective literal $O$ such that $O
\in \cI_T$ and $not(O) \in \cI_F$; $\cI$ is {\em coherent} if $O \in \cI_T$
implies $not(conj_E(O)) \in \cI_F$\footnote{A coherent interpretation
ensures that if some objective literal is explicitly false
(resp. true) then it also must be false (resp. true) by default.}.
The {\em information ordering} of interpretations is defined as
follows. Given two interpretations, $\cI$ and $\cJ$, $\cI
\subseteq_{Info} \cJ$ if $\cI_F$ is a subset of $\cJ_F$, and $\cI_T$
is a subset of $\cJ_T$.  Given an interpretation $\cI$ and a set of
objective literals $\cS$, $\cI|_{\cS}$, the restriction of $\cI$ to
$\cS$, is $\{ L | L \in \cI \mbox{ and } (L \in \cS \mbox{ or } (L =
not(O) \mbox{ and } O \in \cS)) \}$.
%
%---------------------------------------------------------
%
%A consistent three-valued interpretaWtion $M$ is a {\em three-valued
%model} of $P$ if $val_M(C) = {\bf t}$, for every clause $C$ in $P$.
\mycomment{
If $M$ is two-valued then it is called a {\em two-valued\/} or a {\em
total model\/} of $P$.  

Given a ground extended program $P$, we
denote by $WFS(P)$ the (possibly paraconsistent) three-valued
well-founded interpretation of $P$~\cite{VRS91}, extended to include
explicit negation \cite{AlDP95}.
}
%---------------------------------------------------------
% TLS: may be confusing since Interps are both sets and functions.
Any consistent three-valued interpretation can be viewed as a function
from $literals(P)$ to the set $\{{\bf f,u,t}\}$.  Accordingly, for
convenience we assume that the symbols {\bf t} and $not({\bf f})$
belong to every model, while neither {\bf u} nor $not({\bf u})$ belong
to any model.
% where {\bf u}$\equiv not({\bf u}).  
For simplicity of presentation, we assume a left-to-right literal
 selection strategy throughout this paper, although any of the results
 presented here will hold for any fixed literal selection strategy.
 Finally, because dual programs (introduced below) allow any literal
 as the head of a rule, the terms {\em goal}, {\em query} and {\em
 literal} are used interchangeably.

%%%%%%%%%%%%%%%%%%%%%%%%%%%%%%%%%%%%%%%%%%%%%%%%%%%%%%%%%%%%%%%%%%%%%%%%%
\subsection{The Well-Founded Semantics for Extended Programs}
 \label{sec:wfsx} 

We first recall definitions of the well-founded and stable models for
extended programs.  The well-founded model can be seen as a double
iterated fixed point whose inner operators determine a set of true and
false literals at each step.

\begin{definition} \label{def:intops}
For a ground program $P$, interpretation $\cI$ of $P$ and sets $\cO_1$ and
$\cO_2$ of ground objective literals
\begin{itemize} 
\item $Tx^P_{\cI}(\cO_1) = \{O: \mbox{ there is a clause } O \mif{}
L_1,...,L_n \in P \mbox{ and for each } i, 1 \leq i \leq n, L_i \in
\cI \mbox{ or } L_i \in \cO_1 \}$
\item $Fx^P_{\cI}(\cO_2) = \{O: conj_E(O) \in I \mbox{ or (for all
clauses } O \mif{} L_1,...,L_m \in P \mbox{ there exists } i, 1 \leq i
\leq m, conj_D(L_i) \in \cI \mbox{ or } L_i \in \cO_2) \}$
\end{itemize}
\end{definition}

%------------------------------------------------------
\mycomment{
\begin{definition} [{\bf The $\theta_{\cJ}$ Operator}] \label{def:theta-x} 
Let $P$ be an extended program, $\cI$ and $\cJ$ three-valued
interpretations of $P$, and $O$ an objective literal.  The operator
$\theta_{\cJ}$ maps interpretations of $P$ to interpretations of $P$,
and $\theta_{\cJ}(\cI)$ is defined as follows.
\begin{enumerate}
\item $O \in \theta_{\cJ}(\cI)$ iff there is a rule $O \mif L_1,...,L_n$
in $P$ such that for $1 \leq i \leq n$, $L_i$ is an objective literal
and $L_i \in \cI$, or $L_i$ is of the form $not(O)$, and $not(O) \in
\cJ$.
\item $not(O) \in \theta_{\cJ}(\cI)$ iff
\begin{enumerate}
\item for every rule $O \mif L_1,...,L_n$ in $P$, there is a
literal $L_i$ for $1 \leq i \leq n$, such that either $L_i$ is of the
form $not(G)$ and $G \in \cJ$, or $L_i$ is an objective literal and
$not(L_i) \in \cI$; or
\item $conj_E(A) \in \cJ$
\end{enumerate}
\end{enumerate}
\end{definition}
clause 2b, which ensures coherency.
It can be shown \cite{Dama96} that $\theta_{\cJ}$ is monotonic on the
truth ordering for three-valued interpretations, so that it has a
unique least fixed point for a given ${\cJ}$, denoted by
$\omega({\cJ}) = lfp(\theta_{\cJ}(Neg\_Olits))$, where $Neg\_Olits =
\{not(O)| O \in \cH_P\}$.  Furthermore, it can be shown \cite{Dama96}
that the operator $\omega$ is also monotonic on the information
ordering of interpretations leading to the following definition:
%  of the well-founded semantics with explicit negation.
}
%------------------------------------------------------

The only addition required for explicit negation beyond similar
operators for normal programs is the check in the operator
$Fx^P_{\cI}$ that $conj_E(O) \in \cI$, which is used to ensure
coherency.  Both $Tx^P_{\cI}$ and $Fx^P_{\cI}$ can be shown to be
monotonic and continuous over the information ordering by the usual
methods (cf. \cite{Przy89d}), leading to the following operator.

\begin{definition} \label{def:fp_inner}
Let $P$ be a ground program, then $\omega^P_{ext}$ is an operator that
assigns to every interpretation $\cI^1$ of $P$ a new interpretation
$\cI^2$ such that
\[
\begin{array}{l}
	\cI^2_T = lfp(Tx^P_{\cI^1}(\emptyset)) \\
	\cI^2_F = \{ not(O) | O \in gfp(Fx^P_{\cI^1}(objective\_literals(P))) \}
\end{array}
\]
\end{definition}

This latter operation can also be shown to be monotonic over the
information ordering of interpretations by the usual methods, leading
to the formulation of the well-founded semantics as used in this
paper.

\begin{definition} [{\bf Well-founded Semantics for Extended Programs}]
Let $P$ be a ground extended program.  WFS(P) is defined as the least fixed
point, over the information ordering, of $\omega^P_{ext}$.
\end{definition}

\begin{example} \rm
Let $P$ be the program containing the rules $\{ c\mif not(b);\ \ b\mif
a;\ \ -b;\ \ a\mif not(a)\}$.  Then $WFS(P) =
\{-b,c,not(-a),not(b),not(-c)\}$.  Note that to compute $c$, coherency
must be used to infer $not(b)$ from $-b$.
\end{example}

It is important to note that the ``model'' obtained using
$\omega^P_{ext}$ may be paraconsistent.  Using the operator
$\omega^P_{ext}$ it is possible to define a stability operator
%\cite{GeLi88} 
for extended programs that allows partial, and possibly paraconsistent
models.

\begin{definition} [{\bf Partial Stable Interpretation of an Extended
Program}]  \label{def:psi}
Let $P$ be a ground extended program.  We call an interpretation $\cI$ a
{\em partial stable interpretation of P} if $\cI = \omega^P_{ext}(\cI)$
\end{definition}

If an interpretation $\cI$ contains both $O$ and $-O$, then through
coherency, $\omega^P_{ext}(\cI)$ will contain both $O$ and $not(O)$
and so will be inconsistent.  Thus, by definition an interpretation
$\cI$ can be a partial stable interpretation even if it is
inconsistent.  However as we will see, within abductive frameworks
consistency can be ensured by means of integrity constraints --- for
instance, prohibiting $O$ and $-O$ to be true for any objective
literals $O$.
%------------------------------------------------------------
% TLS wording change.
\mycomment{
The reasons $WFSX_P(P)$ is our preferred paraconsistent logic
programming semantics compared to other alternatives are argued at
length in Sections 6 and 8 of \cite{Damp98}.  Besides its desirable
structural properties, it is the only of such semantics where the
support of a literal on a contradiction can be detected by simply
looking at the paraconsistent well-founded model.  } 
%
%------------------------------------------------------------
We use $WFS(P)$ as a basis for abduction in part because the support
of a literal on a contradiction can be detected by simply looking at
the paraconsistent well-founded model.  As shown in Sections 6 and 8
of \cite{Damp98} it is the only one of an array of semantics for
extended programs with this property, along with having other
desirable structural properties.

%------------------------------------------------------------
\mycomment{ TLS: taken out as per reviewer request
Finally, we note that the partial stable interpretations of an
extended program $P$ can also be obtained as the fixed points of
operator $\Gamma(\Gamma_s)$ where $\Gamma_s$ is the application of
$\Gamma$ to the semi-normal version of $P$ \cite{Dama96}.  }
%------------------------------------------------------------

\subsection{Three-Valued Abductive Frameworks}

The definitions of three-valued abductive frameworks modify those of
\cite{DaPe95}.

\begin{definition} [{\bf Integrity Rule}]  \label{def:intrule} An {\em integrity
rule} for a ground program $P$ has the form
\[
\bottom \mif  L_1,\ldots,L_n
\]
where each $L_i$, $1 \leq i \leq n$ is a literal formed over an
element of $\cL_P$. 
\end{definition}

\begin{definition} [{\bf Abductive Framework and Abductive Subgoal}]  
\label{def:abdfrm}
An abductive framework is a triple $< P,\cA,I >$ where $\cA$ is a
finite set of ground objective literals of $\cL_P$ called {\em
abducibles}, such that for any objective literal $O$, $O \in \cA$ iff
$conj_E(O) \in A$, $I$ is a set of ground integrity rules, and $P$ is
a ground program such that (1) there is no rule in $P$ whose head is
in $\cA$; and (2) $\bottom/0$ is a predicate symbol not occurring in
$\cL_P$.

An {\em abductive subgoal} $S = < L,Set >$ is a literal
$L$ together with a finite set of abducibles, $Set$, called the {\em
context} of $S$.  If the context contains both an objective literal
and its explicit conjugate, it is termed {\em inconsistent\/} and is
{\em consistent\/} otherwise.
\end{definition}

Definition~\ref{def:abdfrm} requires that if an objective literal,
say, $-O_1$ is abducible, then $O_1$ must be as well.  This
requirement will be used to allow abduction of positive and negative
information in a symmetric manner.  An abductive subgoal $<L,Set>$
contains a set, $Set$, of such abducibles, along with a subgoal, $L$
which in the dual programs used by \wfsmeth{} can be a literal.  This
notation is used to capture the fact that a solution to $L$ is sought
in the context in which the (positive and negative) objective literals
in $Set$ have been abduced to be true.  If $not(A)$ is a negative
literal, occurring in $P$ or $I$, and $A$ is an abducible objective
literal, \wfsmeth{} will provide coherency axioms to propagate the
truth value of $-A$ or $A$ to $not(A)$ if necessary.  Thus it is
sufficient for the set of abducibles to contain only objective
literals.  The requirement that there can be no rule in $P$ whose head
is an abducible leads to no loss of generality, since any program with
abducibles can be rewritten to obey it \footnote{For instance, if it
is desired to make abducible some objective literal $A$ such that $A$
is the head of a rule, one may introduce a new abducible predicate
$A'$, along with a rule $A \mif{} A'$. See e.g.  \cite{KaKT93}.}.

\begin{definition} [{\bf Abductive Scenario}] \label{def:ab-scen}
A scenario of an abductive framework $< P,\cA, I >$ is a
tuple $< P,\cA,\cB,I >$, where $\cB \subseteq \cA$ is such
that there is no $O \in \cB$, such that $conj_E(O) \in \cB$.
$P_{\cB}$ is defined as the smallest set of rules that contains for
each $A \in \cA$, the rule $A \mif {\bf t}$ iff $A \in \cB$;
%$A \mif {\bf f}$ iff $not(A) \in \cB$; 
and $A \mif {\bf u}$ otherwise.
%if neither $A$ nor $not(A)$ is in $\cB$.
\end{definition}

\begin{definition} [{\bf Abductive Solution}] \label{def:abd-sol}
An abductive solution is a scenario $\sigma = < P,\cA,\cB,I
>$ of an abductive framework, such that $\bottom$ is false in
$M(\sigma) = WFS(P \cup P_{\cB} \cup I)$.
\end{definition}
We say that $\sigma = < P,\cA,\cB,I >$ is an abductive solution for a
query $Q$ if $M(\sigma) \models Q$.  $\sigma$ is minimal, if there is
no other abductive solution $\sigma = < P,\cA,\cB',I >$ for $Q$ such
that $WFS(\cB') \subseteq_{info} WFS(\cB)$.

The definition of an abductive solution is three-valued in that
(objective) literals in $P$, $\cA$, and $I$ may be undefined.  Given a
query and an abductive framework, our goal is to construct a solution
$\sigma$ such that
\[
M(\sigma) \models Q
\]
and
\[
M(\sigma) \models not\ \bot
\]
In addition, it is desirable to evaluate only those portions of $P$
and $I$ that are relevant to $Q$ and to construct solutions that are
minimal in the sense that as few literals as possible are assigned a
value of true or false.  Theorem \ref{thm:exdual} below ensures this
minimality condition.
%
%{\sc Jose: I did not add anything on $\bot$
%being not true: Add something here if you think its important}.
%}
%---------------------------------------------------------

\section{Query Evaluation over Abductive Solutions}
\label{sec:wfs}

We informally introduce \wfsmeth{} through a series of examples
(Formal Definitions can be found in Sections~\ref{sec:unf}
and~\ref{sec:ops}).  \wfsmeth{} shares similarities with SLG in its
propagation of delay literals through \anscr{}, in the semantics it
attaches to unconditional answers, and in its simplification of delay
literals.  The first example illustrates these characteristics.

%-------------------------------------------------------
\begin{example} \rm \label{ex1a}

We first illustrate how \wfsmeth{} can be used to compute queries to
ground programs according to the well-founded semantics when neither
abduction nor integrity constraints are needed.  Accordingly, consider
the abductive framework $< P_1,\emptyset,\emptyset >$, in which the
set of abducibles and the set of integrity rules are both empty, and
$P_1$ is

\begin{CProg}
p \mif  not(q). \\
p \mif  not(r). \\
q \mif  not(p). \\
\end{CProg}

\noindent
$WFS(P_1)$ restricted to the objective literals $\{ p,q,r\}$ is
$\{p,not(q),not(r)\}$.  In order to evaluate the query {\tt ?- q}
through \wfsmeth, we first create the dual form of $P_1$ taken
together with a {\em query rule}
\begin{CProg}
query \mif q, not($\bottom$).
\end{CProg}
where the atom {\tt query} is assumed not to be in $\cL_{P_1}$.  This
rule ensures that integrity constraints are checked for any abductive
solutions that are derived.  This dual program, $dual((\{P_1 \cup query
\mif q,not(\bottom)\}),\emptyset))$ is shown in Figure \ref{fig:P1dual}.

\begin{figure}[htbp]
\longline
\begin{center}
{\tt 
\begin{tabular}{lll}
 & \ \ \ \ \ \ \ &  \\
p :-  not(q).     & \ \ \ \ \ \ \ & not(p) :-  q,r. \\
p :-  not(r).     & \ \ \ \ \ \ \ &  \\
q :-  not(p).     & \ \ \ \ \ \ \ & not(q) :-  p. \\
            & \ \ \ \ \ \ \ &  not(r). \\
query :- q,not($\bottom$).& \ \ \ \ \ \ \ & not(query):- not(q). \\
            & \ \ \ \ \ \ \ & not(query):- $\bottom$. \\
            & \ \ \ \ \ \ \ & not($\bottom$). \\
            & \ \ \ \ \ \ \ & \\
         &not(p) :- -p.  \ \ not(-p) :- p.& \\
         &not(q) :- -q.  \ \ not(-q) :- q.& \\
         &not(r) :- -r.  \ \ not(-r) :- r.& \\
             & \ \ \ \ \ \ \ &
\end{tabular}
}
\end{center}
\longline
\caption{{\em Dual Program for $P_1 \cup \{query \mif q,not(\bottom)\}$}}
\label{fig:P1dual}
\end{figure}

Note that in the dual form of a program, $P$, a rule can have a
default literal of the form $not(A)$ as its head; rules for $not(A)$
are designed to derive $not(A)$ if and only if $A$ is false in
$WFS(P)$.  The last three lines of Figure \ref{fig:P1dual} are {\em
coherency axioms} so-named because they ensure coherency of the model
computed by \wfsmeth{}.  As is usual with tabled evaluations
(e.g. \cite{CheW96}), the \wfsmeth{} evaluation of a query to the
above dual program is represented as a sequence $\cF_0,...,\cF_i$, of
forests of \wfsmeth{} trees.  $\cF_0$ is the forest consisting of the
single tree ${\tt <query,\emptyset> \mif |query}$, which sets up
resolution for the query rule.  Given a successor ordinal $i+1$, a
forest $\cF_{i+1}$ is created when an \wfsmeth{} operation either adds
a new tree to $\cF_i$ or expands a node in an existing tree in
$\cF_i$.  A forest of trees at the end of one possible \wfsmeth{}
evaluation of the above query is shown in Figure
\ref{fig:simpl}~\footnote{For simplicity of presentation, Figure
\ref{fig:simpl} does not display computation paths that include the
coherency axioms, as they are irrelevant in this example.}.  Nodes in
Figure \ref{fig:simpl} are all {\em regular} having the form
$Abductive\_subgoal \mif DelayList|GoalList$, where
$Abductive\_subgoal$ is an abductive subgoal
(Definition~\ref{def:abdfrm}), and $GoalList$ and $DelayList$ are both
sequences of literals.  Intuitively the truth of literals in these
sequences must be determined in order to prove or fail the abductive
subgoal.  When an \wfsmeth{} evaluation encounters a new literal, $S$,
a tree with root $<S,\emptyset> \mif |S$ is added to the forest via
the \newsg{} operation.  Thus, in Figure~\ref{fig:simpl}, when the
literal {\tt q} is selected in node~1, a \newsg{} operation creates
node 2 as a single tree --- indeed, all root nodes other than the
initial node 0 are created through one or another application of this
operation.  Immediate children of the roots of trees are created via
\pgmcr{} operations, while children of other nodes can be created by a
variety of operations to which we now turn.

%------------------------------------------------------------------
\begin{figure}[hbtp]
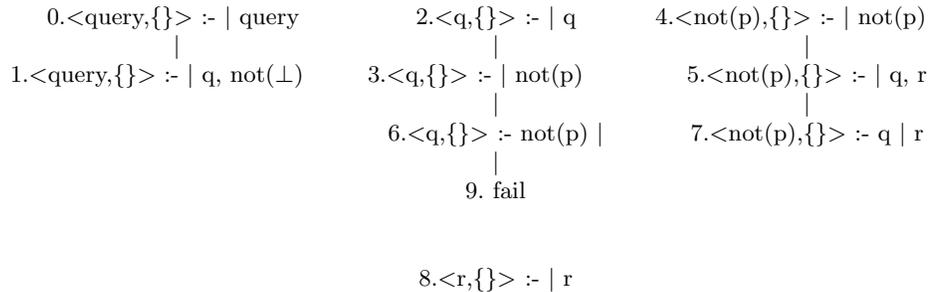

%\centering
%\fbox{\epsfig{file=Figures/simpl.eps,width=\textwidth}},
\longline
\begin{center}
{ 
\begin{tabular}{ccc}
\ \\
0.$<$query,\{\}$>$ :- $|$ query & 2.$<$q,\{\}$>$ :- $|$ q &
4.$<$not(p),\{\}$>$ :- $|$ not(p)\ \ \ \ \   \\

\multicolumn{1}{c}{$|$} & \multicolumn{1}{c}{$|$} & \multicolumn{1}{c}{$|$}\\
1.$<$query,\{\}$>$ :- $|$ q, not($\bottom$)\ \ \ \ \  &
3.$<$q,\{\}$>$ :- $|$ not(p) \ \ \ \ \ &
5.$<$not(p),\{\}$>$ :- $|$ q, r\\
 & \multicolumn{1}{c}{$|$} & \multicolumn{1}{c}{$|$}\\
 & 6.$<$q,\{\}$>$ :- not(p) $|$ & 7.$<$not(p),\{\}$>$ :- q $|$ r\\
 & \multicolumn{1}{c}{$|$} \\
 &\multicolumn{1}{c}{9. fail}\\

& & \\ 
& & \\ 
& 8.$<$r,\{\}$>$ :- $|$ r & \\
& & \\ 
\end{tabular}
}
\end{center}
\longline \caption{Simplified \wfsmeth\  Evaluation of a query to
$<P_1,\emptyset,\emptyset>$} \label{fig:simpl}
\end{figure}
%------------------------------------------------------------------

Consider the state of the evaluation after node 5 has been created.
The evaluation of $q$ depends on $not(p)$ and vice-versa.  In order to
determine the truth of $q$ and $not(p)$ the literal $r$ must be
selected and failed, but this is not possible in a fixed left-to-right
selection strategy.  The \wfsmeth{} \delay{} operation allows the
fixed selection strategy to be broken by moving a selected negative
literal from the $GoalList$ of a node to its $DelayList$ so that
further literals in the node, such as $r$, may be selected.  Applied
to node 3, the \delay{} operation produces node 6, ${\tt <q,\emptyset>
\mif{} not(p) |}$.  An {\em answer} is a regular leaf node with an
empty $GoalList$.  In the subforest of Figure \ref{fig:simpl}
consisting of nodes whose index is 6 or less, node 6 is an answer.
Because its $DelayLists$ is non-empty it is termed a {\em conditional
answer}.  While node 6 is an answer for $q$, it is not known at the
time node 6 is created whether $q$ is true or false --- its truth
value is conditional on that of $not(p)$.  Answers are returned to
other nodes via the \anscr{} operation which also combines the
abductive contexts of the answer and the node to ensure consistency.
Using this operation, the conditional answer is resolved against the
selected literal of node 5 producing node 7, ${\tt <not(p),\emptyset>
\mif{} q | r}$.  Similarly to SLG, the \anscr{} operation of
\wfsmeth{} does not propagate $DelayLists$ of conditional answers,
thus the literal added to the $DelayList$ in node 7 is $q$ rather than
the literal originally delayed, $not(p)$.  This action is necessary
for \wfsmeth{} to have polynomial complexity for normal programs in
the absence of abduction (cf. Theorem \ref{thm:complex}).  If
$DelayLists$ of conditional answers were propagated directly, the
number of answers for a given subgoal could be proportional to the
number of its derivations (see \cite{CheW96} for an example of such a
program).  Thus a literal $L$ can be added to the $DelayList$ of a
node in one of two ways: if $L$ is negative, it can be added through
an explicit \delay{} operation; otherwise, $L$ can be added to a
$DelayList$ if an \anscr{} operation resolves a conditional answer
against $L$ regardless of whether $L$ is positive or negative.

Note that after the production of node 8, the evaluation of $not(p)$
and of all the selected subgoals in the goal list upon which it
depends cannot proceed further, and these subgoals cannot produce any
new answers, conditional or otherwise.  Such subgoals are termed {\em
completely evaluated} (Definition~\ref{def:compl-eval}).  At this
stage, node 6 contains in its $DelayList$ an atom that is known to be
false -- i.e. that is completely evaluated and has no answers.  A
\simpl{} operation is applicable to node 6, creating the failure node,
node 8, as its child, so that node 6 is no longer a leaf and hence no
longer an answer. After the production of node 9, neither the tree for
$q$ nor that for $not(p)$ has an answer at the end of the evaluation,
corresponding to the fact that both literals are false in $WFS(P_1)$.
\end{example}

%-------------------------------------------------------

We now formalize the definitions of some concepts introduced in
Example \ref{ex1a}.  For an objective literal $O$ in a program $P$,
$not(O)$ is defined so that it will be derivable as true iff $O$ is
false in $WFS(P)$.  For instance, if there is a fact in $P$ for some
objective literal $O$ then the dual has no rule for $not(O)$.  The
definition below is somewhat more complicated than the form implicitly
used in Example~\ref{ex1a}, but as explained below, it ensures both
that \wfsmeth{} will be definable on infinite programs and that it
will have an appropriate complexity for finite programs.

%-------------------------------------------------------
\begin{definition} [{\bf Dual Program}] \label{dual-fold}
Let $P$ be a ground extended program, and $\cA$ a (possibly empty)
finite set of abducibles.  The {\em dual transformation} creates a
{\em dual program} $dual(P,\cA)$, defined as the union of $P$ with
smallest program containing the sets of rules $fold_P$ and $cohere_P$
as follows:
\begin{enumerate}
\item Let $O$ be an objective literal for which there are no facts
in $P$, and with $\beta \leq \omega$ rules of the form:
\[
\begin{array}{rrl}
r_i: &	 	O\mif & L_{i,1},...,L_{i,n_1}\\ 
\end{array}
\]
for $i < \omega$, where each $n_i$ is finite.
\begin{enumerate}
\item Then $fold_P$ contains the rule
\[
\begin{array}{rl}
	not(O)\mif & not(fold^a_1\_O)
\end{array}
\]
along with rules
\[
\begin{array}{rl}
	not(fold^a_i\_O) \mif & not(fold^b_i\_O),not(fold^a_{i+1}\_O)
\end{array}
\]
for all $i$, $1 \leq i < \beta$; and
\[
\begin{array}{rl}
	not(fold^a_{\beta}\_O) \mif not(fold^b_{\beta}\_O)& 
\end{array}
\]
if $\beta$ is finite.
\item 
and for $1 \leq j \leq maximum\{n_1,...,n_{i}\}$, such that $L_{i,j}$
exists as a literal in $r_i$, $fold_P$ contains a rule:
\[
\begin{array}{rl}
	not(fold^b_i\_O) \mif & conj_D(L_{i,j}) \\
\end{array}
\]
\end{enumerate}
where $fold^a_k\_O$, $fold^b_k\_O$ are assumed not to occur in $\cL_P$
for any $k$ (such rules are termed {\em folding rules}, and literals
formed from objective literals whose predicate symbol is $fold^a_k\_O$
or $fold^b_k\_O$ are called {\em folding literals}).
\item Otherwise, if $not(O)$ is in $literals(P)$, but there is no rule
with head $O$ in $P$, then $fold_P$ contains the rule $not(O) \mif {\bf
t}$.  If there is a fact for $O$ in $P$, the rule $not(O) \mif {\bf f}$
may be introduced or omitted.
\item $cohere_P$ consists of {\em axioms of coherence} that relate explicit
and default negation, defined as:
\[
not(O) \mif conj_E(O)
\]
For each objective literal $not(O)$ in either $literals(P \cup fold_P)$
or $\cA$.
\end{enumerate}
\end{definition}

\begin{example} \rm
Consider a program fragment in which an objective literal {\tt m} is
defined as:
\begin{CProg}
m \mif  n$_1$, not(o$_1$).\\
m \mif  n$_2$, not(o$_2$).\\
m \mif  n$_3$, not(o$_3$).\\
\end{CProg}
Note that a naive dualization of {\tt m} as implicitly used in
Example~\ref{ex1a} (and as defined in Definition~\ref{dual-unfold})
would produce a rule for each partial truth assignment to the body
literals of {\tt m} that falsifies {\tt m}, leading to 8 rules, each
with 3 body literals.  Indeed, it is easy to see that naive
dualization of a predicate $p$ with $\beta$ clauses can lead to a
predicate for $not(p)$ that has a number of clauses exponential in
$\beta$, making the naive dual form unsuitable in terms of complexity
for finite programs.  Furthermore, the number of body literals in a
clause for $not(p)$ may be linear in $\beta$ so that if the naive
transformation were used, the dual of an infinite program would not be
a program as defined in Section~\ref{sec:terminology}.

The folding rules in the dual form of $m$ ($fold_P$ of
Definition~\ref{dual-fold}) are shown in Figure~\ref{fig:folded-dual}.
%------------------------------------------------------------------
\begin{figure}[htbp]
\longline
\begin{center}
{\tt {\small
\begin{CProg}
\\
not(m) \mif not(fold$^a_1$\_m).\\
\\
not(fold$^a_1$\_m) \mif not(fold$^b_1$\_m),not(fold$^a_2$\_m). \\
not(fold$^a_2$\_m) \mif not(fold$^b_2$\_m),not(fold$^a_3$\_m). \\
not(fold$^a_3$\_m) \mif not(fold$^b_3$\_m). \\
\\
not(fold$^b_1$\_m):- not(n$_1$).\\
not(fold$^b_1$\_m):- o$_1$. \\
not(fold$^b_2$\_m):- not(n$_2$).\\
not(fold$^b_2$\_m):- o$_2$. \\
not(fold$^b_3$\_m):- not(n$_3$).\\
not(fold$^b_3$\_m):- o$_3$.  \\
\ 
\end{CProg}
}}
\end{center}
\longline
\caption{{\em Folded Dual Program for a program clause}} 
\label{fig:folded-dual}
\end{figure}
%------------------------------------------------------------------
In Definition~\ref{dual-fold}, if there are an infinite number of
rules defining an objective literal $O$, there will also be an
infinite number of folding rules defining $not(O)$, but each rule will
have a finite sequence of literals in their body.  Also note that in a
finite ground program, if an objective literal $O$ is defined by $n$
rules each of which have $m$ body literals, the size of the rules
defining $O$ will be $mn+n$ (see Definition~\ref{def:size} for a
precise definition of the size of rules and programs).  In
$dual(P,\cA)$, there will be $m*n$ rules of the form $fold^b\_O_i$ for
some $i$, each of size 2, along with folding rules of the type
$fold^a\_O_i$ for some $i$ so that the size of the rules for $not(O)$
in $dual(P,\cA)$ is linear in the size of the rules for $O$ in $P$.
\end{example}

While the dual form of Definition~\ref{dual-fold} is necessary for the
correctness and complexity results that follow, examples will use a
simpler form without folding literals that is logically equivalent for
finite programs (see Definition~\ref{dual-unfold} for an exact
statement of this simpler form).

%-------------------------------------------------------

\begin{definition} [{\bf \wfsmeth{} Trees and Forest}]
\label{def:slg-forest}  
An \wfsmeth{} forest consists of a forest of \wfsmeth{} trees.  Nodes
of \wfsmeth{} trees are either {\em failure nodes} of the form {\em fail},
or {\em regular nodes} of the form
\[
Abductive\_Subgoal \mif DelayList | GoalList
\]
where $Abductive\_subgoal$ is an abductive subgoal
(Definition~\ref{def:abdfrm}).  Both $DelayList$ and $GoalList$ are
finite sequences of literals (also called {\em delay literals} and
{\em goal literals}, respectively).

We call a regular leaf node $N$ an {\em answer} when
\emph{GoalList} is empty.  If {\em DelayList} is also empty, $N$ is
{\em unconditional}; otherwise it is {\em conditional}.
\end{definition}

Definition~\ref{def:ops} will ensure that the root node of a given
\wfsmeth{} tree, $T$, has the form $< S, \emptyset > \mif
| S$, where $S$ is a literal.  In this case, we say that $S$ is the
{\em root goal} for $T$ or that $T$ is the {\em tree for} $S$.
Similarly by Definition~\ref{def:ops}, a forest contains a root goal
$S$ if the forest contains a tree for $S$.  Literal selection rules
apply to the $GoalList$ of a node; as mentioned in
Section~\ref{sec:prelim}, we use a fixed left-to-right order for
simplicity of presentation so that the leftmost literal in the
$GoalList$ of a node is termed the {\em selected literal} of the node.

%----------------------------------------------------------
\begin{example} \rm \label{ex1}

The well-founded semantics captures infinite recursion by means of the
concept of unfounded sets: an atom involved in an unfounded set is
assigned a truth-value of false.  When a program undergoes the dual
transformation, negative literals involved in infinite recursion must
be made to succeed.  As an example of this, consider the abductive
framework $<P_2,\emptyset,\emptyset>$ in which $P_2$ is defined as:

\begin{CProg}
s \mif  not(p), not(q), not(r). \\
p \mif  not(s), not(r), q. \\
q \mif  not(p), r. \\
r \mif  not(q), p.
\end{CProg}

\noindent
Note that $WFS(P_2)$ restricted to the objective literals ${\{
s,p,q,r}\}$ is $\{ s, not(p), not(q), not(r)\}$.  Assuming the query
{\tt ?- s} to $<P_2,\emptyset,\emptyset>$, the dual program
$dual((\{P_2 \cup query \mif s,not(\bottom)\}),\emptyset)$ is shown in
Figure~\ref{fig:P2dual}~\footnote{The transformation in
Definition~\ref{dual-unfold} is used for simplicity.}

\begin{figure}[htbp]
\longline
\begin{center}
{\tt
\begin{tabular}{l@{}ll}
        & \ \ &  \\
s:-  not(p),not(q),not(r).
        & \ \ & not(s):-  p. \\
            & \ \ & not(s):-  q. \\
            & \ \ & not(s):-  r. \\
p:-  not(s),not(r),q.
        & \ \ & not(p):-  s. \\
            & \ \ & not(p):-  r. \\
            & \ \ & not(p):-  not(q). \\
q:-  not(p),r.
        & \ \ & not(q):-  p. \\
            & \ \ & not(q):-  not(r). \\
r:-  not(q),p.
        & \ \ & not(r):-  q.  \\
            & \ \ & not(r):-  not(p).  \\
query:- s,not($\bottom$).
        & \ \ & not(query):- not(s). \\
            & \ \ & not(query):- $\bottom$. \\
            & \ \ & not($\bottom$). \\
            & \ \ & \\
         &not(p):- -p.  \ \ not(-p):- p.& \\
         &not(q):- -q.  \ \ not(-q):- q.& \\
         &not(r):- -r.  \ \ not(-r):- r.& \\
         &not(s):- -s.  \ \ not(-s):- s.& \\
             & \ \ \ \ \ \ \ &
\end{tabular}
}
\end{center}
\longline
\caption{{\em Dual Program for $P_2 \cup \{query \mif s,not(\bottom)\}$}}
\label{fig:P2dual}
\end{figure}

%------------------------------------------------------------------
\begin{figure}[hbtp]
\longline \vspace{10pt}
%\centering
%    \fbox{\epsfig{file=Figures/wfs.eps,width=\textwidth}}
\footnotesize

\[\begin{array}{l@{\ \ \ \ \ \ \ \ \ \ \ \ \ \ \ \ \ \ \ \ \ }r}
 \fbox{
  \begin{array}[t]{c}
  $0.$<$query,\{\}$>$ :- $|$ query\tiny$\\
  |\\
  $1.$<$query,\{\}$>$ :- $|$ s, not($\bottom$)$\\
  |\\
  $33.$<$query,\{\}$>$ :- $|$ not($\bottom$)$\\
  |\\
  $36.$<$query,\{\}$>$ :- $|\\
 \end{array}
 }
 &
  \fbox{
  \begin{array}[t]{c}
  $2.$<$s,\{\}$>$ :- $|$ s$\\
  |\\
  $3.$<$s,\{\}$>$ :- $|$ not(p), not(q), not(r)$\\
  |\\
  $30.$<$s,\{\}$>$ :- $|$ not(q), not(r)$\\
  |\\
  $31.$<$s,\{\}$>$ :- $|$ not(r)$\\
  |\\
  $32.$<$s,\{\}$>$ :- $| \
 \end{array}
 }
\end{array}\]

 $ $\\ $ $\\

\[ \fbox{ \begin{array}[t]{c@{}c@{}c}
  \multicolumn{3}{c}{\!\!\!\!\!\!\!\!\!\!$4.$<$not(p),\{\}$>$ :- $|$ not(p)$}\\
\multicolumn{3}{l}{$\begin{picture}(315,10)
\put(140,10){\line(0,-1){10}}
\put(140,10){\line(-6,-1){65}} \put(140,10){\line(6,-1){65}} \end{picture}$}\\
  \begin{array}[t]{c}
   \multicolumn{1}{r}{$5.$<$not(p),\{\}$>$ :- $|$ s$}\\
   |\\
   $37.$<$not(p),\{\}$>$ :- $| \\
  \end{array}
  &
  \begin{array}[t]{c}
   \ \ $6.$<$not(p),\{\}$>$ :- $|$ r$\\
  \end{array}
  &
  \begin{array}[t]{c}
   $23.$<$not(p),\{\}$>$ :- $|$ not(q)$\\
   |\\
   $24.$<$not(p),\{\}$>$ :- not(q) $| \\
   |\\
   $27.$<$not(p),\{\}$>$ :- $| \\
  \end{array}
 \end{array}}\]

 $ $\\ $ $\\

\[\begin{array}{lr}
 \fbox{ \begin{array}[t]{c@{}c}
 \multicolumn{2}{l}{$\ \ \ \ 9.$<$not(q),\{\}$>$ :- $|$ not(q)$}\\
\multicolumn{2}{l}{$\begin{picture}(200,10)
\put(70,10){\line(0,-1){10}}
\put(70,10){\line(6,-1){65}} \end{picture}$}\\
\begin{array}[t]{c}
   $10.$<$not(q),\{\}$>$ :- $|$ p$\\
  \end{array}
  &
  \begin{array}[t]{c}
   $19.$<$not(q),\{\}$>$ :- $|$ not(r)$\\
   |\\
   $25.$<$not(q),\{\}$>$ :- not(r) $|\\
   |\\
   $28.$<$not(q),\{\}$>$ :- $|\\
  \end{array}
 \end{array}}&
 \fbox{
 \begin{array}[t]{c}
  $7.$<$r,\{\}$>$ :- $|$ r$\\
  |\\
  $8.$<$r,\{\}$>$ :- $|$ not(q), p$\\
  |\\
  $38.$<$r,\{\}$>$ :- $|$ p$\\
 \end{array}
 }
 \end{array}\]

 $ $\\ $ $\\

 \[\fbox{ \begin{array}[t]{c@{}c@{}c}
  \multicolumn{3}{c}{$13.$<$not(s),\{\}$>$ :- $|$ not(s)$}\\
%  \multicolumn{2}{c}{\ \ \ \ \ \ \ \ \ \ /\ \ \ \ \ $|$} & \multicolumn{1}{l}{$\backslash$}\\
\multicolumn{3}{l}{$\begin{picture}(315,10)
\put(140,10){\line(0,-1){10}} \put(140,10){\line(-6,-1){65}}
\put(140,10){\line(6,-1){65}} \end{picture}$}\\
\begin{array}[t]{c}
   \multicolumn{1}{r}{$14.$<$not(s),\{\}$>$ :- $|$ p$}\\
  \end{array}
  &
  \begin{array}[t]{c}
   $15.$<$not(s),\{\}$>$ :- $|$ q$\\
  \end{array}
  &
  \begin{array}[t]{c}
   $18.$<$not(s),\{\}$>$ :- $|$ r$\\
  \end{array}
 \end{array}}\]

 $ $\\ $ $\\

\[\begin{array}{lr}
\fbox{\begin{array}[t]{c@{}c}
  \multicolumn{2}{l}{$\ \ \ \ 20.$<$not(r),\{\}$>$ :- $|$ not(r)$}\\
\multicolumn{2}{l}{$\begin{picture}(200,10)
\put(70,10){\line(0,-1){10}}
\put(70,10){\line(6,-1){65}} \end{picture}$}\\
  \begin{array}[t]{c}
   $21.$<$not(r),\{\}$>$ :- $|$ q$\\
  \end{array}
  &
  \begin{array}[t]{c}
   $22.$<$not(r),\{\}$>$ :- $|$ not(p)$\\
   |\\
   $26.$<$not(r),\{\}$>$ :- not(p) $|\\
   |\\
   $29.$<$not(r),\{\}$>$ :- $|\\
  \end{array}
 \end{array}}
& \fbox{ \begin{array}[t]{c}
  $16.$<$q,\{\}$>$ :- $|$ q$\\
  |\\
  $17.$<$q,\{\}$>$ :- $|$ not(p), r$\\
  |\\
  $41.$<$q,\{\}$>$ :- $|$ r$\\
 \end{array}}
\end{array}\]

\[\begin{array}{l@{\ \ \ \ \ \ \ \ \ \ \ \ \ \ \ \ \ \ \ \ \ \ \ \ \ \ \ \ }r}
 \fbox{ \begin{array}[t]{c}
  $11.$<$p,\{\}$>$ :- $|$ p$\\
  |\\
  $12.$<$p,\{\}$>$ :- $|$ not(s), not(r), q$\\
  |\\
  $39.$<$p,\{\}$>$ :- not(s) $|$ not(r), q$\\
  |\\
  $40.$<$p,\{\}$>$ :- not(s) $|$ q$\\
 \end{array}}
  &
 \fbox{ \begin{array}[t]{c}
  $34.$<$not($\bottom$),\{\}$>$ :- $|$ not($\bottom$)$\\
  |\\
  $35.$<$not($\bottom$),\{\}$>$ :- $|\\
 \end{array}}
\end{array}\]

\vspace{10pt} \longline

\normalsize \caption{{\em Simplified \wfsmeth\  Evaluation of a
query to $<P_2,\emptyset,\emptyset>$}} \label{fig:wfs}
\end{figure}
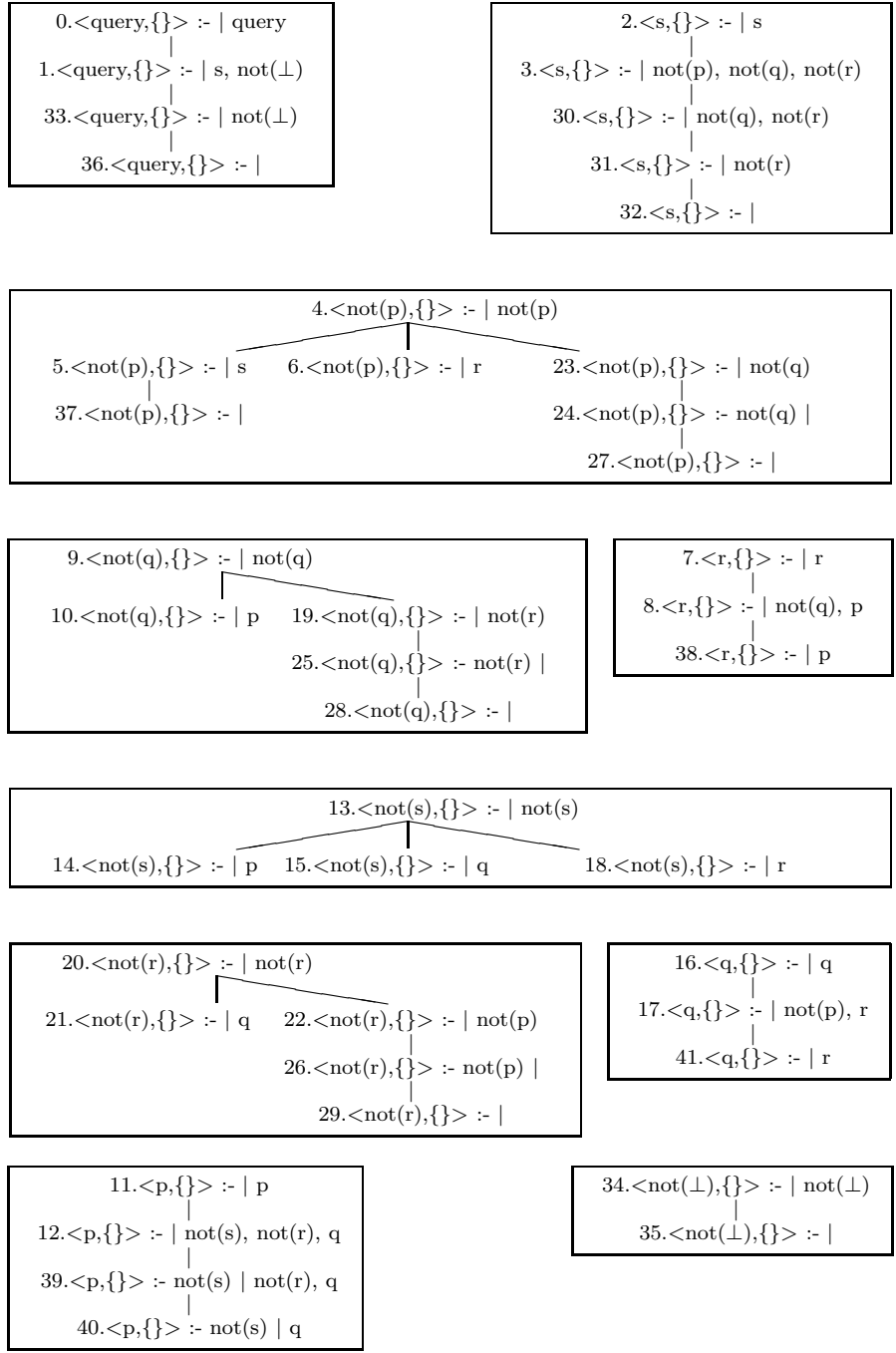
%------------------------------------------------------------------

An \wfsmeth{} forest at the end of an evaluation of $query$ is shown
in Figure~\ref{fig:wfs}.  As can be seen from Figure~\ref{fig:wfs},
the evaluation at first proceeds using the same operations as in
Example~\ref{ex1a}, where the roots of non-initial trees are created
via \newsg{} operations, the children of roots of trees created via
\pgmcr{} operations, and other nodes created via \anscr{} or \delay{}
operations.  However node 27 is produced by a new operation.  Note
that in the subforest of Figure~\ref{fig:wfs} consisting of nodes
numbered 26 or less, nodes 24, 25, and 26 are all conditional answers
that ``depend'' on each other through their $DelayList$s.  However, in
the well-founded model of $P_2$, $p$, $q$ and $r$ should be false as
they belong to an unfounded set (based on the empty interpretation).
In order to derive their truth-values \wfsmeth{} includes a \unfr{}
operation.  Nodes 24, 25, and 26 together form an analogue in the dual
program to an unfounded set \cite{VRS91} consisting of $p$, $q$, and
$r$ in $P_2$.  Such an analogue is called a {\em co-unfounded set}.
Whereas positive literals in an unfounded set are all false, negative
literals in a co-unfounded set are all true.  When an answer is
determined to belong to a co-unfounded set, it is made unconditionally
true.  In this example the \unfr{} operation creates the unconditional
answer, node 27, while \simpl{} operations produce nodes 28 and 29.
Subsequently, \anscr{} resolves the (unconditional) answer {\tt
$<$not(p),$\emptyset> \mif |$} against the selected literal of node 3 to
create node 30 through \anscr{}, and subsequent applications of this
operation produce nodes 31-34, 36-38, 40 and 41.
\end{example}

%----------------------------------------------------------
%
We summarize some of the elements of the previous two examples.
Intuitively, the distinction between goal literals and delay literals
is that goal literals are currently selected within a node or are yet
to be selected.  As a result, there is an answer for $S$ if there is a
regular leaf node $N$ in a tree for $S$ that has no goal literals.  If
$N$ does not contain delay literals, it is an unconditional answer and
$S$ has an abductive solution defined by the context of the abductive
subgoal of $N$; if $N$ does contain delay literals, then it is a
conditional answer and the abductive context for $N$ is not yet
determined by the evaluation to make $N$ either true or false.
Finally, a completely evaluated subgoal
(Definition~\ref{def:compl-eval}) that has no answers at all is
interpreted to be false for all abductive contexts.  At an operational
level, as described in Definition~\ref{def:ops}, goal literals may be
resolved away via an \anscr{} operation, abduced, or delayed.
Literals in the $DelayList$ were not resolved away when they were
selected, but rather their resolution was postponed.  Delay literals
are subject to the \unfr{} operation mentioned in Example~\ref{ex1},
and also to \simpl{} operations mentioned in Example~\ref{ex1a}.
Maintaining both delay literals and goal literals within an evaluation
is useful as it is necessary to identify unfounded sets of objective
literals within the well-founded semantics, as well as co-unfounded
sets of objective literals within the dual form of a program.
Determining (co-~)unfounded sets is expensive in practical terms, so
that restricting such an operation to delay literals can form an
important optimization (cf. \cite{DiSw02}).

The notion of a set of \wfsmeth{} trees being completely evaluated was
introduced in Example~\ref{ex1a} to capture the concept of when a set
of trees in a forest has returned all of the answers in the model of a
program.  This can happen in one of two ways.  First, a tree may
contain an unconditional answer whose abductive context is empty, in
which case further evaluation will not produce any more minimal
abductive answers.  Second, a tree may have had all possible
\wfsmeth{} operations performed on the selected literal in the
$GoalList$ of each of its nodes.  For this condition to occur, all
possible answers must have been returned to the selected literals so
that a tree is not completely evaluated unless all trees that it
depends on (through the selected literal of each of its nodes) are
completely evaluated as well.  An example of this occurs in
Example~\ref{ex1}, where the trees for $p$, $q$, and $r$ are mutually
dependent and may only be evaluated together.

\begin{definition} [{\bf Completely Evaluated}]\label{def:compl-eval}
Given an \wfsmeth{} forest $\cF$, a set ${\cal T}$ of \wfsmeth{} trees
is \emph{completely evaluated} iff at least one of the following
conditions is satisfied for each tree $T \in {\cal T}$:

\begin{enumerate}
\item $T$ contains an unconditional answer whose abductive subgoal
context is empty; or 
%\vspace{-.2cm}
\item For each node $N$ in ${\cal T}$ with selected goal literal $SL$ 
\begin{itemize}
%\vspace{-.2cm}
\item The tree for $SL$ belongs to a set $S'$ of completely evaluated
trees; and
\item No \newsg, \pgmcr, \anscr{}, \delay{}, or {\sc Abduction}
operations (Definition~\ref{def:ops}) are applicable to $N$.
\end{itemize}
\end{enumerate}
A literal $L$ is completely evaluated in $\cF$ if the tree for $L$
belongs to a completely evaluated set in $\cF$.
\end{definition}

Finally, we turn to an example to illustrate how \wfsmeth{} can
evaluate queries to general abductive frameworks.

%----------------------------------------------------------
\begin{example} \rm \label{ex2}
Consider the abductive framework $< P_3,\cA_3,I_3 >$, in which $P_3$
is the program
\begin{CProg}
p\mif {\tt not(q*)}. \\
q\mif {\tt not(p*)}.
\end{CProg}
$\cA_3 = \{p*,q*,-p*,-q*\}$, and $I_3$ is the program
\begin{CProg}
$\bottom$\mif p\_constr \\
$\bottom$\mif q\_constr \\
p\_{constr}\mif p, -p*. \\
q\_{constr}\mif q, -q*.
\end{CProg}
So that the (ground) integrity constraints represent an abductive
interpretation of default negation.  Let the query rule be
\begin{CProg}
query \mif q,not($\bottom$).
\end{CProg}
The dual program with coherency axioms (simplified for presentation by
using the transformation of Definition~\ref{dual-unfold}, which does
not include folding predicates.) is shown in Figure~\ref{fig:p3dual}.
\begin{figure}[hbtp]
\longline
\begin{center}
{\tt {\small
\begin{tabular}{lll}
            & \ \ \ \ \ \ \ &  \\
p\mif not(q*).  & \ \ \ \ \ \ \ & not(p)\mif q*.    \\
q\mif not(p*).  & \ \ \ \ \ \ \ & not(q)\mif p*.    \\
$\bottom$\mif p\_constr & \ \ \ \ \ \ \ & not($\bottom$)\mif
                    not(p\_constr),not(q\_constr)  \\
$\bottom$\mif q\_constr     & \\
p\_constr\mif p, -p*.    & \ \ \ \ \ \ \ & not(p\_{constr})\mif not(p) \\
            & \ \ \ \ \ \ \ & not(p\_{constr})\mif not(-p*). \\
q\_constr\mif q, -q*.&  \ \ \ \ \ \ \ & not(q\_{constr})\mif not(q) \\
            & \ \ \ \ \ \ \ & not(q\_{constr})\mif not(-q*). \\
query\mif q,not($\bottom$).& \ \ \ \ \ \ \ & not(query)\mif not(q). \\
            & \ \ \ \ \ \ \ & not(query)\mif $\bottom$. \\
            & \ \ \ \ \ \ \ & \\
not(-p) \mif p.         & \ \ \ \ \ \ \ & not(p) \mif -p. \\
not(-q) \mif q      & \ \ \ \ \ \ \ & not(q) \mif -q \\
not(-p*) \mif p*    & \ \ \ \ \ \ \ & not(p*) \mif -p* \\
not(-q*) \mif q*    & \ \ \ \ \ \ \ & not(q*) \mif -q* \\
not(-p\_constr) \mif p\_constr  & \ \ \ \ \ \ \ & not(p\_constr) \mif -p\_constr \\
not(-q\_constr) \mif q\_{constr} & \ \ \ \ \ \ \ & not(q\_constr) \mif -q\_{constr} \\
            & \ \ \ \ \ \ \ &  \\
\end{tabular}
}}
\end{center}
\longline
\caption{{\em Dual program for $P_3 \cup \{ query \mif q,not(\bottom)\} \cup \cA \cup I$}}
\label{fig:p3dual}
\end{figure}

\begin{figure}[htbp]
\longline \vspace{10pt}
%\centering \mbox{
%\fbox{\epsfig{file=Figures/abex1.eps,width=\textwidth}}}
% %{\epsfig{file=slg-forest-sss.eps,height=10.4cm}}}
\footnotesize

\[\begin{array}{l@{\ \ \ \ \ \ \ \ \ \ \ \ \ \ \ \ \ \ \ \ \ \ \ \ \ \ \ \ \ }r}
 \fbox{ \begin{array}[t]{c}
  $0.$<$query,\{\}$>$ :- $|$ query$\\
  |\\
  $1.$<$query,\{\}$>$ :- $|$ q, not($\bottom$)$\\
  |\\
  $5.$<$query,\{-p*\}$>$ :- $|$ not($\bottom$)$\\
  |\\
  $30.$<$query,\{-p*q*\}$>$ :- $| \\
 \end{array}}
 &
 \fbox{ \begin{array}[t]{c}
  $2.$<$q,\{\}$>$ :- $|$ q$\\
  |\\
  $3.$<$q,\{\}$>$ :- $|$ not(p*)$\\
  |\\
  $4.$<$q,\{-p*\}$>$ :- $|\\
 \end{array}}
\end{array}\]

\[\fbox{\begin{array}[t]{c@{}c}
  \multicolumn{2}{c}{$6.$<$not($\bottom$),\{p*\}$>$ :- $|$ not($\bottom$)$}\\
  \multicolumn{2}{c}{|}\\
  \multicolumn{2}{c}{$7.$<$not($\bottom$),\{p*\}$>$ :- $|$ not(p\_constr), not(q\_constr)$}\\
  \multicolumn{2}{l}{$\begin{picture}(200,10) \put(140,10){\line(-4,-1){45}} \put(190,10){\line(4,-1){45}}
\end{picture}$}\\
  \begin{array}[t]{l}
   \multicolumn{1}{c}{$11.$<$not($\bottom$),\{p*\}$>$ :- $|$ not(q\_constr)$}\\
   \multicolumn{1}{l}{$\begin{picture}(100,10) \put(100,10){\line(-4,-1){45}} \put(100,10){\line(1,-2){13}}
 \end{picture}$}\\
   $15.$<$not($\bottom$),\{p*,q*\}$>$ :- $| \\ \\ \multicolumn{1}{r}{$21.$<$not($\bottom$),\{p*\}$>$ :- $|}\\
  \end{array} &
  \begin{array}[t]{l}
   \multicolumn{1}{c}{$27.$<$not($\bottom$),\{q*\}$>$ :- $|$ not(q\_constr)$}\\
   \multicolumn{1}{l}{$\begin{picture}(100,10) \put(100,10){\line(-4,-1){45}} \put(100,10){\line(1,-2){13}}
 \end{picture}$}\\
   $28.$<$not($\bottom$),\{q*,p*\}$>$ :- $| \\ \\ \multicolumn{1}{r}{$29.$<$not($\bottom$),\{q*\}$>$ :- $|}\\
  \end{array}
 \end{array}}\]

\[
  \fbox{ \begin{array}[t]{cc}
   \multicolumn{2}{c}{$8.$<$not(p\_constr),\{\}$>$ :- $|$
not(p\_constr)$}\\
   \multicolumn{2}{l}{$\begin{picture}(287,10)
\put(130,10){\line(-4,-1){45}} \put(140,10){\line(4,-1){45}}
\end{picture}$}\\
   $9.$<$not(p\_constr),\{\}$>$ :- $|$ not(-p*)$&
 $22.$<$not(p\_constr),\{\}$>$ :- $|$ not(p)$\\
   | & |\\
   $10.$<$not(p\_constr),\{p*\}$>$ :- $| &
  $26.$<$not(p\_constr),\{q*\}$>$ :- $|\\
  \end{array}}
 \]

\[\fbox{ \begin{array}[t]{cc}
   \multicolumn{2}{c}{$12.$<$not(q\_constr),\{\}$>$ :- $|$ not(q\_constr)$}\\
   \multicolumn{2}{l}{$\begin{picture}(287,10)
\put(130,10){\line(-4,-1){45}} \put(140,10){\line(4,-1){45}}
\end{picture}$}\\
   $13.$<$not(q\_constr),\{\}$>$ :- $|$ not(-q*)$& $16.$<$not(q\_constr),\{\}$>$ :- $|$ not(q)$\\
   | & |\\
   $14.$<$not(q\_constr),\{q*\}$>$ :- $| &
$20.$<$not(q\_constr),\{p*\}$>$ :- $|\\
  \end{array}}
\]

\[\begin{array}{l@{\ \ \ \ \ \ \ \ \ \ \ \ \ \ \ \ \ \ \ \ \ \ \ \ \ \ \ }r}
  \fbox{\begin{array}[t]{c}
   $17.$<$not(q),\{\}$>$ :- $|$ not(q)$\\
   |\\
   $18.$<$not(q),\{\}$>$ :- p*$|\\
   |\\
   $19.$<$not(q),\{p*\}$>$ :- $|\\
  \end{array}}
 &
  \fbox{\begin{array}[t]{c}
   $23.$<$not(p),\{\}$>$ :- $|$ not(p)$\\
   |\\
   $24.$<$not(p),\{\}$>$ :- q*$|\\
   |\\
   $25.$<$not(p),\{q*\}$>$ :- $|\\
  \end{array}}\\
\end{array}\]

 \vspace{10pt} \longline

\normalsize
\caption{{\em Simplified \wfsmeth{} evaluation of a
query to $<P_3,\cA_3,I_3>$}.} \label{fig:abex1}
\end{figure}
\end{example}

%Note that no separate evaluation method is needed for integrity
%constraints --- rather, the call to $not(\bottom)$ directly calls rules in
%the dual program.
Figure \ref{fig:abex1} illustrates a forest of trees created by an
\wfsmeth{} evaluation of this initial query.  For purposes of space,
it does not depict derivations stemming from coherency axioms.  When
an abductive framework contains a non-trivial set of abducibles,
provision must be made for when the selected literal of a given node
is an abducible, as well as for propagating abducibles among abductive
subgoals.  In the first case, if the selected literal of a node $N$ is
an abducible, and the addition of the selected literal to the context
of the abductive subgoal of $N$ does not make the context inconsistent
(Definition \ref{def:abdfrm}), an {\sc Abduction} operation is
applicable to $N$.  For instance, {\sc Abduction} operations are used
to produce nodes 25 and 19.  The figure also illustrates cases in
which abducibles are propagated through \anscr{}.  Node 5 is produced
by resolving the answer {\tt <q,\{-p*\}> \mif |} against the selected
literal, $q$ of node 1, to produce the (consistent) context $\{-p*\}$.
Abducibles therefore differ from delay literals in that the abducibles
are propagated into the context of an abductive subgoal, while delayed
literals are not propagated into delay lists.  Propagating abducibles
through \anscr{} operations is common in this derivation, producing in
a similar manner, nodes 20, 4, 11, 27, 15, 21, 28, 29, 9, 26, and 14.

Certain of these nodes are created using the coherency axioms, which
are not shown in Figure~\ref{fig:abex1}.  For instance in producing
node 4, {\tt <q,\{-p*\}> :- |}, a \newsg{} operation creates a new
tree for the selected literal, $not(p*)$ of node 3.  This tree uses
the rule {\tt not(p*) \mif{} -p*} for \pgmcr{}, and then abduces {\tt
-p*}, propagating the abducible to the context of node 4.  In
propagating abducibles, the \anscr{} operation enforces the
restriction that the context of the answer must be consistent with the
context of the abductive subgoal of the node to which the answer is
returned.  For instance, of the two unique abductive solutions to {\tt
not($\bottom$)} only one can be returned to the node {\tt
$<$query,\{-p*\}$>$\mif | not($\bottom$)}, namely {\tt
$<$not($\bottom$),\{q*\}}$>$.

%----------------------------------------------------------

The final definitions for \wfsmeth{} are now provided, beginning with
the unfounded and co-unfounded sets.

%---------------------------------------------------------
\subsection{Unfounded and Co-unfounded Sets} \label{sec:unf}

One of the ideas behind of the well-founded semantics of normal
programs is to assign the value of false to atoms that are contained
in unfounded sets.  Intuitively these sets can be seen as including
atoms whose derivations lead to positive loops or to infinite chains
of dependencies among subgoals.  Unfounded sets for extended logic
programs are defined as follows:

\begin{definition} [{\bf Unfounded Set of Objective Literals}]
\label{def:unfounded} 
Let $P$ be a ground extended logic program, and $\cI$ a coherent
interpretation of $P$.  Then a set of objective literals $\cS
\subseteq literals(P)$ is an unfounded set of $P$ with respect to
$\cI$ if for each rule $r_s$ with head $H \in \cS$ one of the
following conditions hold:
\begin{enumerate}
\item for some body literal $L_i$ in $r_s$, the default conjugate of
$L_i$ is in $\cI$.
\item for some positive body literal $L_i$ in $r_s$, $L_i
\in \cS$.
\end{enumerate}
A literal that makes either condition true is called a {\em witness of
unusability} for rule $r_s$ with respect to $\cI$.
\end{definition}

A witness of unusability in $\cI$ may be a literal that is false in
$\cI$, or a positive literal whose proof depends on positive literals
that are neither contained in $\cI$ nor are provable from literals in
$\cI$.  For instance, in the program $\{p \mif{} q, q \mif{} p\}$ both
$p$ and $q$ are unfounded when $\cI =\emptyset$.  In a dual program,
there is also the dual notion of a {\em co-unfounded set of literals}.

\begin{definition} [{\bf Co-unfounded set of literals}]
\label{def:co-unfounded-lits} 
Let $P$ be a ground extended program, $\cA$ a set of abducibles, and
$\cI$ a coherent interpretation of $dual(P,\cA)$.  Then a set of
negative literals $\cS \subseteq literals(dual((P,\cA))$ is a
co-unfounded set with respect to $\cI$ if for each $H \in \cS$, there
is a rule $H \mif{} Body$ such that for each $L_i
\in Body$:
\begin{enumerate}
\item $L_i$ is true in $\cI$; or
\item $L_i \in \cS$.
\end{enumerate}
\end{definition}

Just as unfounded sets of objective literals are false in a program,
co-unfounded sets of negative literals are true in the dual of a
program (cf. Lemma~\ref{lem:co-unfounded-wfs}).  Because any selected
negative literal can be delayed, \wfsmeth{} need only take account of
co-unfounded sets of literals that occur in $DelayLists$ of nodes.
For instance, evaluation of the program in Example~\ref{ex1} required
detection of a co-unfounded set among literals in the $DelayLists$ of
nodes 24,25, and 26.  A {\em co-unfounded set of answers} corresponds
to a co-unfounded set of literals that arises in a
\wfsmeth{} evaluation, and is defined as:

%-----------------------------------------------------------------
% TLS: original definition did not specify negative literal.
% TLS: changed to ensure that each literal is in the $DelayLists$.
% TLS: changed $S$ to be non-empty.
\begin{definition} [{\bf Co-unfounded Set of Answers}] \rm
\label{def:co-unfounded} 
Let $\cF$ be an \wfsmeth{} forest, and $\cS$ a non-empty set of answers
in $\cF$.  Then $\cS$ is a {\em co-unfounded set} in $\cF$ iff
\begin{enumerate}
\item Each literal $S_i$, such that $<S_i,C_i>$ is the abductive
subgoal of an answer in $\cS$, is a completely evaluated negative
literal.  Further, $S_i$ is contained in the $DelayList$ of some answer
in $\cS$.
\item The set 
\[
Context = \bigcup \{ C_i | <S_i,C_i> \mif DL |  \mbox{ is an answer in } \cS \}
\]
is consistent; and 
\item For each answer $<S_i,C_i> \mif DL_i| \in \cS$ 
\begin{enumerate}
\item $DL_i$ is non-empty; and 
\item for each $S_j \in DL_i$, there exists an answer $<S_j,Context_j> \mif
DL_j| \in \cS$.
%\item there is a path in the delay dependency graph of $\cF$ from each
%$S_j \in DL$ to $S_i$ 
\end{enumerate}
\end{enumerate}
\vspace{-.1in}
\end{definition}
%-----------------------------------------------------------------

The requirement in condition 1 of Definition~\ref{def:co-unfounded}
that the literals be completely evaluated is for convenience, so that
an evaluation need not detect co-unfounded sets of answers when more
direct derivations may still be possible.  

%-------------------------------------------------------
Analogous to a co-unfounded set of answers are the non-supported
objective literals.  Intuitively, non-supported literals in an
\wfsmeth{} forest correspond to unfounded objective literals
under a given interpretation.

\begin{definition} [{\bf Supported Objective Literals}] \label{def:support}
\label{def:ans-types} \rm 
Let $\cF$ be a forest, and $S$ a positive literal that is the root
goal for a tree $T$ in $\cF$.  Then $S$ is supported in $\cF$ iff
\begin{enumerate}
\item $T$ is not completely evaluated; or
\item $T$ contains an answer $<S,Context> \mif DL|$ in $T$ with no
positive delay literals in $DL$; or
\item $T$ contains an answer $<S,Context> \mif DL|$ in $T$ such
      that, every positive delay literal~$L_1$ in~$DL$ is supported in
      $\cF$.
\end{enumerate}
\end{definition}

A tree in a forest is thus supported if it is not completely
evaluated, if it contains an unconditional answer, if it contains an
answer with a delayed negative literal, or if it contains an answer
containing positive literals all of which are themselves supported. 
The \simpl{} operation of Definition~\ref{def:ops} removes an answer
of an unfounded literal from a forest by creating a failure node as a
child of the answer.

%-------------------------------------------------------
\subsection{\wfsmeth{} Evaluations and Operations} \label{sec:ops}

An \wfsmeth{} evaluation consists of a (possibly transfinite) sequence
of \wfsmeth{} forests \footnote{Our definition here follows that of
\cite{Swif99b} for generalized SLG trees.}.  In order to define the
behavior of an \wfsmeth{} evaluation at a limit ordinal, we define a
notion of a least upper bound for a set of \wfsmeth{} trees.  
\mycomment{ TLS: not needed If a
global ordering on objective literals is assumed, then the elements in
the $Context$ of the abductive subgoal of a node can be uniformly
ordered, and using this ordering an equivalence relation can be
defined for nodes of \wfsmeth{} trees with equivalent abductive
contexts.  }
Any rooted tree can be viewed as a partially ordered set in which each
node $N$ is represented as $\{N,P\}$ in which $P$ is a tuple
representing the path from $N$ to the root of the tree.  When
represented in this manner, it is easily seen that when $T_1$ and
$T_2$ are rooted trees, $T_1 \subseteq T_2$ iff $T_1$ is a subtree of
$T_2$, and furthermore, that if $T_1$ and $T_2$ have the same root,
their union can be defined as their set union, for $T_1$ and $T_2$
taken as sets.  However, we will sometimes abuse notation in our
definitions and refer to trees using the usual graph-theoretic
terminology.

%-------------------------------------------------------
\begin{definition} [{\bf \wfsmeth{} Evaluation}] \label{def:wfsmeth-eval}
Let $< P,\cA,I >$ be an abductive framework and $Q$ a
query.  An \wfsmeth{} evaluation $\cE$ of $Q$ to $< P,\cA,I
>$ is a sequence of \wfsmeth{} forests $\cF_0,\cF_1,...,\cF_n$
operating on the ground instantiation of $dual((P \cup I \cup \{query
\mif not(\bottom),Q\}),\cA)$ such that:
\begin{itemize}
\item $\cF_0$ is the forest containing the single tree,
$<query,\emptyset> \mif | query $,
\item For each successor ordinal $n+1$, $\cF_{n+1}$ is
obtained from $\cF_n$ by applying an \wfsmeth{} operation from
Definition~\ref{def:ops}.
\item For each limit ordinal $\alpha$, $\cF_{\alpha}$ is
defined such that $T \in \cF_{\alpha}$ iff 
\begin{itemize}
\item The root node of $T$, $<S,\emptyset> \mif |S$ is the root node of
some tree in a forest $F_i$, $i < \alpha$;
\item $T = \cup_{i < \alpha}(\{T_i | T_i \in \cF_i \mbox{ and } T_i
\mbox{ has root } <S,\emptyset> \mif |S)$
\end{itemize}
\end{itemize} 
If no operation is applicable in $\cF_n$, then it is called a
\emph{final forest} of $\cal{E}$.  
\end{definition}
%-------------------------------------------------------

In accordance with Definition~\ref{def:wfsmeth-eval}, the following
\wfsmeth{} operations operate on dual programs.

\begin{definition} [{\bf \wfsmeth{} Operations}] \label{def:ops}
Let $\cF_n$ be an \wfsmeth{} forest for an evaluation of a query $Q$
to an abductive framework $< P,\cA,I >$, and suppose $n+1$
is a successor ordinal.  Then $\cF_{n+1}$ may be produced by one of
the following operations

\begin{enumerate}
\item  \newsg:
Let $\cF_n$ contain a non-root node 
\[
N = <S,Context> \mif DL|L, GoalList.
\]
If $L$ is not an abducible and $\cF_n$ contains no tree with root goal
$L$, add the tree: 
$ <L,\emptyset> \mif |L.  $
\item \pgmcr:
Let $\cF_n$ contain a root node 
\[
N = <S,\emptyset> \mif | S
\] and let there be a clause $S \mif Body$ in the dual program.  If in
$\cF_n$, $N$ does not have a child: 
\[
N_{child} = <S,\emptyset> \mif | Body
\]
then add $N_{child}$ as a child of $N$.
\item \anscr:
Let $\cF_n$ contain a non-root node 
\[ 
N = <S,Context_1> \mif DL_0 | L,Body 
\]
and suppose that $\cF_n$ contains an answer node $<L,Context_2>\mif
DL_1|$, such that $Context_1 \cup Context_2$ is consistent.  Let $DL_2
= DL_0,L$\ \ if $DL_1$ is not empty, and $DL_2 = DL_0$ otherwise.
Finally, if in $\cF_n$, $N$ does not have a child
\[
N_{child} = <S,Context_1 \cup Context_2> \mif DL_2 | Body
\]
then add $N_{child}$ as a child of $N$.
\item \delay{}: Let $\cF_n$ contain a non-root leaf node 
\[
N = <S,Context>\mif DL| not(L),Body
\]
where $L$ is not an abducible, and where $\cF_n$ contains a tree for
$not(L)$, but no answer of the form $<not(L),\emptyset> \mif|$.
%, but $not(G)$ is not completely evaluated.  
Then add: \ 
$
	<S,Context>\mif DL,not(L) | Body
$
\ as a child of $N$.
\item \simpl :  Let $N =  <S,Context_1> \mif DL|$ be
	a node for a tree with root goal $S$, and let $D$ be a delay
 	literal in~$DL$.  Then
\begin{itemize} 
\item if $\cF_n$ contains an unconditional answer node  $<D,Context_2> \mif
 	|$, and if $Context_1 \cup Context_2$ is consistent, let $DL_1
 	= DL - D$. 	If
\[
N_{child} =  	<S,Context_1 \cup Context_2> \mif DL_1| 
\]
	is not a descendant of $N$ in $\cF$, add $N_{child}$ as a
	child of $N$. 
\item if the tree for $D$ is completely evaluated and contains no
 	answers whose context is consistent with $C_1$; or if $D$ is a
 	positive literal that is non-supported, then create a child
 	$fail$ of $N$.
\end{itemize}
\item \unfr{}:  Let
\[
N =  <S,Context_S> \mif DL| 
\] 
be an answer in $\cF_n$, such that there is a minimal co-unfounded set
of answers $\cS$ in $\cF_n$ containing $N$ together with answers
$<L_i,Context_i> \mif DL_i |$ for all literals $L_i \in DL$.  Let
\[
Context_{union} = Context_S \cup  \bigcup_{<L_i,Context_i> \mif
DL_i \in \cS} Context_i 
\]
Then if $N$ does not have a child
$
N_{child} = <S,C_{union}> \mif |
$,
create a child $N_{child}$ of $N$.
\mycomment{
\item \unfr{}:  Let $\cS$ be a co-unfounded set of answers and
let $C_{union}$ be the consistent union of all contexts of answers in
$\cS$.  Then for each
\[
N =  <S,C_1> \mif DL| \in S
\] 
create a child of $N$:
$
N_{child} = <S,C_{union}> \mif |
$
}
\item {\sc Abduction}: Let 
\[
N = <S, Context>\mif DL|A,Body
\]
where $A$ is an abducible
\mycomment{ or its default negation,}
and suppose that $\{ A \} \cup Context$ is consistent.  Finally,
assume that in $\cF_n$, $N$ does not have a child
\[
N_{child} = <S,Context \cup \{ A \}>\mif DL|Body
\]
Then add $N_{child}$ as a child of $N$.
\end{enumerate}
\end{definition}

For a discussion of the similarities between definitions of \wfsmeth{}
and those of SLG see Section \ref{sec:disc}.
%-------------------------------------------------------
\subsection{Soundness and Completeness of \wfsmeth{}} \label{sec:snc}

The first result on the correctness of \wfsmeth{} concerns the
correctness of the dual transformation itself
(Definition~\ref{dual-fold}).  To show this, we introduce fixed point
operators for dual programs that are analogous to those of
Section~\ref{sec:wfsx}, and show that they can be used to construct
the well-founded semantics.  As before, there are two sets of
operators.  The first operators form the inner fixed point for dual
programs, and are analogous to the operators of
Definition~\ref{def:intops}.

\begin{definition} \label{def:dualintops}
For a ground program $P$ and set $\cA$ of abducibles, let
$dual(P,\cA)$ be the dual program formed by applying the dual
transformation to $P$ and $\cA$.  Let $\cO_1$ be a set of ground
objective literals formed over $dual(P,\cA)$ and $\cL_1$ be a set of
ground negative literals formed over $dual(P,\cA)$.
\begin{itemize} 
\item $Td^{dual(P,\cA)}_{\cI}(\cO_1) = \{O | O \mbox{ is an objective literal
and } O \mif{} L_1,...,L_n \in dual(P,\cA)
\mbox{ and for each } 1 \leq i \leq n, L_i \in \cI \mbox{ or } L_i \in
\cO_1 \}$
\item $Fd^{dual(P,\cA)}_{\cI}(\cL_1) = \{not(O) |  O \mbox{ is an
objective literal and } not(O) \mif{} L_1,...,L_n
\in dual(P,\cA) \mbox{ and for each } 1 \leq i \leq n, L_i \in \cI \mbox{
or } L_i \in \cL_1 \}$
\end{itemize}
\end{definition}

The operator $Td^{dual(P,\cA)}_{\cI}$ is essentially the same as the
operator $Tx^{P}_{\cI}$ of Definition~\ref{def:intops}; however the
operator $Fd^{dual(P,\cA)}_{\cI}$ differs significantly from its
analogue, since negative literals are defined by clauses in the dual
program that are to be made true.  As before, these operators can be
shown to be monotonic by the usual methods, leading to the following
operator which is used in the outer fixed point.

\begin{definition} \label{def:dualfp_inner}
For a ground program $P$ and set $\cA$ of abducibles, let
$dual(P,\cA)$ be the dual program formed by applying the dual
transformation to $P$ and $\cA$.  Let
\[
\cS_{dual(P,\cA)} = \{not(O) | not(O) \in literals(dual(P,\cA)) \} 
\]
$\omega_d^{dual(P,\cA)}$ is an operator that assigns to every
interpretation $\cI^1$ of $P$ a new interpretation $\cI^2$ such that
\[
\begin{array}{l}
	\cI^2_T = lfp(Td^P_{\cI^1}(\emptyset)) \\
	\cI^2_F = gfp(Fd^P_{\cI^1}(\cS_{dual(P,\cA)}))
\end{array}
\]
\end{definition}

In the above definition, the use of the greatest fixed point of the
operator $Fd^P_{\cI^1}(\cS_{dual(P,\cA)})$ captures the fact that
co-unfounded sets need to be made true when evaluating the dual
program.  The following theorem shows the correctness of the dual
transformation.

\begin{theorem} \label{thm:dual-equiv}
For a ground program $P$ and empty set of abducibles, let
$dual(P,\emptyset)$ be the dual program formed by applying the dual
transformation to $P$.  Then for any literal $L \in literals(P)$,
\[
L \in WFS(P) \iff L \in lfp(\omega_d^{dual(P,\emptyset)})
\]
where the least fixed point of $lfp(\omega_d^{dual(P,\emptyset)})$ is
taken with regard to the information ordering of interpretations.
\end{theorem}

Recall that for a query $Q$ to an abductive framework, an \wfsmeth{}
evaluation as stated in Definition~\ref{def:wfsmeth-eval} represents
solutions to $Q$ that are consistent with the integrity rules by means
of a query rule $query \mif{} Q, not(\bottom)$.  The following theorem
shows that the \wfsmeth{} operations, acting on the dual program
correctly compute abductive solutions.

\begin{theorem} \label{thm:exdual}
Let $< P, \cA, I >$ be an abductive framework, and $\cE$
be an \wfsmeth{} evaluation of $Q$ against $< P, \cA, I
>$.  Then
\begin{itemize}
\item $\cE$ will have a final forest $\cE_{\beta}$;
\item if $<query,Set>\mif|$ is an answer in $\cE_\beta$, $\sigma
= < P,A,Set,I >$ is  an abductive solution for $< P, \cA, I >$;
\item if $< P,A,Set,I >$ is a minimal abductive solution
for $Q$, then $<query,Set>\mif|$ is an answer in
$\cE_\beta$.
\end{itemize}
\end{theorem}
\begin{proof}
The proof is contained in the Appendix.
\end{proof}

%-------------------------------------------------------

\subsection{Finite Termination and Complexity of \wfsmeth{} for 
%Evaluating
Extended Programs} \label{sec:term-compl}

Termination of \wfsmeth{} evaluations is guaranteed under the
following conditions.

\begin{theorem} \label{termination}
Let $< P,\cA,I >$ be an abductive framework such that $P$
and $I$ are finite ground extended programs, and $\cA$ is a finite set
of abducibles.  Let $\cE$ be an \wfsmeth{} evaluation of a query $Q$
against $< P,\cA,I >$.  Then $\cE$ will reach a final
forest after a finite number of \wfsmeth{} operations.
\end{theorem}
\begin{proof}
The proof is contained in the Appendix.
\end{proof}

It is known that the problem of query evaluation to abductive
frameworks is NP-complete, even for those frameworks in which
entailment is based on the well-founded semantics~\cite{EiGo96}.  More
precise results can be obtained for \wfsmeth, as shown in the
following theorem, which uses a summation over abductive contexts
(using of the combinatorial selection function ``choose'') to
determine the cost of an \wfsmeth{} evaluation.  The following theorem
relies on a definition of size that is made precise in
Definition~\ref{def:size}.

\begin{theorem} \label{thm:complex}
Let $\cF$ be the final forest in an \wfsmeth{} evaluation $\cE$ of a
query $Q$ against a finite ground abductive framework $< P, \cA,
I >$.  Let $C_{context}$ be the maximal cardinality of the
context of any abductive subgoal in $\cF$, and $C_{abducibles}$ be the
cardinality of $\cA$.  Then $\cF$ can be constructed in $M \times 2
\times size(dual((P \cup I),\cA))$ steps, where
\[
M = \sum_{i \leq C_{context}} \left( \begin{array}{c}
					C_{abducibles} \\ i
				     \end{array}
			      \right)
\]
\end{theorem}
\begin{proof} 
The proof is given in the Appendix.
\end{proof}

Intuitively, this theorem states that the complexity of an \wfsmeth{}
evaluation is proportional to the maximal number of abducibles in any
abductive subgoals, and to the number of abducibles in the framework.
If the number of either of these factors can be reduced, then the
complexity of the evaluation will be reduced.  Since the size of a
dual program is linear in the size of an abductive framework
(cf. Lemma~\ref{dual-sizes}), a corollary of Theorem
\ref{thm:complex} is that if the set of abducibles and integrity rules
are both empty, the final forest of a \wfsmeth{} evaluation requires a
number of operations that is linear in the size of the input program.
It is important to note, however, that \wfsmeth{} operations may not
be implementable with constant cost.  In particular, some operations
such as \unfr{} or removal of a non-supported answers may require a
cost that is linear in the size of a program so that the cost of
evaluating a program with empty abducibles and integrity rules may not
be linear but will remain polynomial (see \cite{DiSw02} for an
extended discussion of costs of tabling normal programs).

Theorem \ref{thm:complex} can be used to show that abduction over the
well-founded semantics is {\em fixed-parameter tractable}
\cite{DoFe95}.  Recall that a decision problem $Pr$ can be defined as a
question to be answered with a ``yes'' or ``no''.  This question has
several input parameters.  If particular values of these input
parameters $I$ are given, an {\em instance} of the problem, $Pr_I$ is
given.  Informally, a {\em parameterized decision problem} $Pr(k)$
(and its instances) can be defined by designating a certain input
parameter $k$ so that its complexity is an explicit function of this
parameter.  With this background, the following definition is adapted
from \cite{GoSS99}.

\begin{definition}
Let $Pr(k)_I$ be an instance of a parameterized decision problem
$Pr(k)$ and $Pr_I$ the instance of the non-parameterized version of
$Pr(k)$.  Then $Pr(k)$ is {\em (strongly uniformly) fixed-parameter
tractable} if there is an algorithm that decides whether $Pr(k)_I$ is
a yes-instance of $Pr(k)$ in time $f(k_s)\cO(n^c)$ where $n$ is the
size of $Pr_I$, $k_s$ is an integer parameter, $c$ is a constant and
$f$ a recursive function.
\end{definition}

In order to show that abduction over the well-founded semantics is
fixed-parameter tractable, consider the decision problem of whether
$Q$ is contained in the abductive solution (Definition
\ref{def:abd-sol}) to a finite, ground abductive framework, $<
P, A, I >$.  Then, since $C_{context}$ of
Theorem~\ref{thm:complex} is not greater than $C_{abducibles}$, $k_s$
can be set to $C_{abducibles}$ and $f(k_s)$ the summation, $M$, in
Theorem~\ref{thm:complex} with $C_{context}$ replaced by
$C_{abducibles}$.  Furthermore, $size(dual((P \cup I),\cA))$ is linear
in the size of $P \cup A \cup I$ by Lemma~\ref{dual-sizes}, so that
when the maximal size of an abductive context is factored out,
evaluation of $Q$ requires a number of \wfsmeth{} operations linear in
the size of $P \cup A \cup I$.

The above considerations lead to the following Theorem.

\begin{theorem}
Let $Q$ be a query to a finite ground abductive framework $< P,
\cA, I >$, and let $C_{abducibles}$ be the size of the
set $\cA$.  Then the problem of deciding whether $Q$ is contained in
an abductive solution to $< P, A, I >$ is fixed-parameter
tractable with respect to $C_{abducibles}$.
\end{theorem}

There are, of course other means for parameterizing abduction over the
well-founded semantics.  For instance, an estimate of the maximum
cardinality of contexts of abductive subgoals could be made via a
suitably defined dependency graph, so that the input parameter
$C_{abducibles}$ could be replaced by this parameter.  Alternately,
\wfsmeth{} might be adapted so that a restriction were placed on the
size of all abductive contexts.  Such an approach might be relevant to
using \wfsmeth{} to solve model-based diagnosis problem, where
attention was restricted to identifying single or double faults in the
model.

%-------------------------------------------------------

\section{Construction of Generalized Stable Models through
\wfsmeth} \label{sec:gsm}

The three-valued abductive frameworks of Section \ref{sec:prelim} are
not the only semantics used for abduction: Generalized Stable Models
\cite{KaMa90} provide an important alternative.  In \cite{DaPe95} it was
shown that the 
%three-valued 
abductive framework of Section \ref{sec:prelim} has the same
expressive power as generalized stable models.
%\cite{KaKT93}.  
In this section, we reformulate these results to show that \wfsmeth{}
can be used to evaluate abductive queries over generalized stable
models.  By allowing all positive literals to be inferred through
abduction, \wfsmeth{} can be used to construct partial stable
interpretations (Definition \ref{def:psi}).  By choosing appropriate
integrity constraints, these interpretations can be constrained to be
consistent and total.  We begin by adapting the concept of a
generalized stable model to the terminology of
Section~\ref{sec:prelim}.

\begin{definition} [{\bf Generalized Partial Stable Interpretation and Model}] 
\label{def:gsm} 
Let $< P,\cA,I >$ be an abductive framework, with a
scenario $\sigma = < P,\cA,\cB,I >$. Then $M(\sigma)$ is a
generalized partial stable interpretation of $< P,\cA,I >$
if
\begin{itemize}
\item  $M(\sigma)$ is a partial stable interpretation of $< P
\cup P_{\cB}  \cup I >$; and 
\item $\bot$ is false in $M(\sigma)$.
\end{itemize}
If in addition $M(\sigma)$ is an answer set of $< P \cup P_{\cB}
\cup I >$, $\sigma$ is a generalized stable model of $< P,\cA,I
>$. 
\end{definition}

Generalized stable models can be computed by adding additional program
rules, abducibles, and integrity constraints to abductive frameworks
and computing the solution to these frameworks as per Definition
\ref{def:abd-sol}.

\begin{definition} \label{def:gsmtransform}
Let $< P,\cA,I >$ be an abductive framework.  Then let
$\cS$ be the smallest set containing a new objective literal,
$abd\_O$, not in $literals(P \cup I \cup \cA)$ for each objective
literal $O$ in literals $literals(P \cup I \cup \cA)$.  A literal
formed over an element of $\cS$ is called a {\em shadow literal}.
Let
\[
R = O \mif Body
\]
be a rule in $P \cup I$.  Then a {\em shadow rule} for $R$ is a rule 
\[
R_{shadow} = O \mif Body_{abd}
\]
in which each literal of the form $not(O')$ in $Body$ is replaced by
$not(abd\_O'$).  The shadow rules are denoted $Shadow(P)$ for a
program $P$.  Corresponding to these shadow rules are {\em shadow
constraints} ($I_{shadow}$) of the form
\[
\begin{array}{rl}
\bot\mif& O, not(abd\_O). \\
\bot\mif& not(O), abd\_O.
\end{array}
\]
for each $abd\_O$ such that $not(abd\_O) \in literals(Shadow(P))$.

The {\em consistency constraints} ($I_{consist}$) for $< P,\cA,I
>$  consist of the shadow constraints along with integrity rules
of the form 
\[
\begin{array}{rl}
\bot\mif& O,not(O).  \\
%\bot\mif& O,-O. \\
% TLS took these two out.
% \bot\mif& O, not(abd\_O). \\
% \bot\mif& not(O), abd\_O.
\end{array}
\]
for $O \in literals(P \cup I)$.

The {\em totality rules} ($I_{total}$) for $< P,\cA,I >$ have the form
\[
\begin{array}{rl}
	\bot\mif & not(defined_O) \\
	defined_O\mif & O\\ 
	defined_O\mif & not(O)\\ 
\end{array}
\]
for each $O \in literals(P \cup I)$.
\end{definition}

\begin{example} \rm \label{ex:gsms}
Consider the abductive framework consisting of the program $P_3$: 
\begin{CProg}
p\mif {\tt not q}. \\ 
q\mif {\tt not p}.
\end{CProg}
with an empty set of abducibles and integrity constraints.  In order
to compute the partial stable interpretations of $P_3$ via abductive
solutions, shadow rules must be added along with integrity and
consistency constraints.  For simplicity, we ignore coherency rules
below.  The shadow rules of $P$, $Shadow(P)$ are
\begin{CProg}
p\mif {\tt not abd\_q}. \\ 
q\mif {\tt not abd\_p}.
\end{CProg}
While the shadow constraints, $I_{shadow}$, include the rules
\begin{CProg}
$\bot$\mif p, not abd\_p\\ 
$\bot$\mif q, not abd\_q\\ 
$\bot$\mif not p, abd\_p\\ 
$\bot$\mif not q, abd\_q\\ 
\end{CProg}
and the consistency constraints, $I_{consist}$, include all
instantiations of the schemata 
\begin{CProg}
$\bot$\mif O, not O\\ 
$\bot$\mif O, -O\\ 
\end{CProg}
for $O \in literals(P \cup I)$.  Let $\cA$ be the set $\{abd\_O \mbox{
or } -abd\_O | not(abd\_O) \in literals(Shadow(P) \}$.  Then the
abductive framework
\[
< (P \cup Shadow(P)), A, (I_{shadow} \cup I_{consist}) >:
\]
has solutions
\[
\begin{array}{l}
\sigma_1 = \{ abd\_q , abd\_p \}	\\
\sigma_2 = \{ -abd\_q  \}	\\
\sigma_3 = \{ -abd\_p \}
\end{array}
\]
These solutions correspond to the following generalized partial stable
interpretations of $< P,\emptyset,\emptyset >$, whose
restrictions to the atoms of $P$ are:
\[
\begin{array}{l}
M(\sigma_1)|_{ \{p,q \}}  = \emptyset	\\
M(\sigma_1)|_{ \{p,q\}}  = \{ p \}	\\
M(\sigma_1)|_{ \{p,q\}}  = \{ q \}
\end{array}
\]
Note that, in accordance with the definitions of
Section~\ref{sec:prelim}, positive and negative objective literals are
abduced, and coherency propagates negation from abduced objective
literals to negative literals.  In order to derive the generalized
stable models of $< P,\emptyset,\emptyset >$, the totality
constraints of Definition~\ref{def:gsmtransform} must also be added.
In the above example, the totality constraints would prevent the first
scenario, $\sigma_1$, from being an abductive solution.
\end{example}

\mycomment{
\[
\begin{array}{rl}
	\{abd\_p,abd\_q\} \\
	\{p, -abd\_q\} \\
	\{q,-abd\_p\} 
\end{array}
\]
}
Example \ref{ex:gsms} illustrates the following theorem.

\begin{theorem} \label{thm:gsm}
Let $F = < P, \cA, I >$ be an abductive framework, and
$\sigma = < P,A,\cB,I >$ be an abductive scenario for $F$.
Let $Shadow(P \cup I)$ be the set of shadow rules for $P \cup I$, and
let $I_{shadow}$, $I_{consist}$, and $I_{total}$ be the shadow,
consistency, and totality constraints for $F$ as in
Definition~\ref{def:gsmtransform}.  Let $\cA_{shadow} =
\{abd\_O | not(abd\_O) \in Shadow(P \cup I)\}$.  Then
\begin{enumerate}
\item $M(\sigma)$ is a generalized partial stable interpretation of
$< P,A,I >$ iff there exists an abductive solution
\[
\sigma' = < (P \cup Shadow(P \cup I)),(A \cup A_{shadow}),\cB,(I \cup
I_{shadow}) > 
\]
such that $M(\sigma) = M(\sigma')$.
\item $M(\sigma)$ is a generalized stable model of $< P,A,I
>$ iff there exists an abductive solution 
\[ 
\sigma' = < (P \cup Shadow(P \cup I)),(A \cup A_{shadow}),\cB,(I \cup I_{shadow} \cup
I_{consist} \cup I_{total} > 
\]
such that $M(\sigma) = M(\sigma')$.
\end{enumerate}
\end{theorem}
\begin{proof}
This result is straightforward from Definition \ref{def:gsm} and the
results of \cite{DaPe95}.
\end{proof}

Theorem~\ref{thm:gsm} has several implications.  First, since the
paraconsistent well-founded model of a program is a partial stable
interpretation, use of the shadow program and constraints includes
computation of the paraconsistent well-founded model as a special
case.  In addition, because Theorem~\ref{thm:exdual} states that
\wfsmeth{} can be used for query evaluation to abductive frameworks
based on $WFS$, \wfsmeth{} can be used to compute queries to
generalized partial stable interpretations and generalized stable
models.  The cost of this computation, of course, includes the cost of
potentially evaluating shadow rules and the various additional
integrity constraints.  It is known that the problem of deciding the
answer to a ground query to an abductive framework is NP-complete when
the entailment method is based on the well-founded semantics
\cite{EiGo96}, as is the problem of deciding whether an abductive
framework has a generalized stable model.  The lack of polynomial data
complexity of \wfsmeth{} for arbitrary abductive frameworks is
therefore understandable, given the power of these frameworks.
Finally, we note that computation of consistent answer sets can also
be obtained via the transformation in
Definition~\ref{def:gsmtransform}.

\section{Discussion} \label{sec:disc}

\subsection{A Meta-interpreter for \wfsmeth{} and its Applications}

Currently the \wfsmeth{} system is implemented on top of the XSB
System~\cite{XSB-home}.  It consists of a preprocessor for generating
the dual program, plus a meta-interpreter for the tabled evaluation of
abductive goals, and is available from {\tt
http://www.cs.sunysb.edu/\symbol{126}tswift}. This meta-interpreter
has the termination property of Theorem~\ref{termination}, but does
not have the complexity property of Theorem \ref{thm:complex}.  Work
is currently being done in order to migrate into the XSB engine some
of the tabling mechanisms of \wfsmeth{} now taken care by the
meta-interpreter, such as the \unfr{} operation.

\paragraph*{Psychiatric Diagnosis}  \wfsmeth{} was originally
motivated by a desire to implement psychiatric diagnosis
\cite{GSTPD00}. Knowledge about psychiatric disorders is codified by
{\em DSM-IV} \cite{DSM-IV} sponsored by the American Psychiatric
Association.  Knowledge in DSM-IV can be represented as a directed
graph with positive links to represent relations from diagnoses to
sub-diagnoses or to symptoms.  These graphs also have negative links,
called {\em exclusion} links that represent symptoms or diagnoses that
must shown false in order to derive the diagnosis.  The DSM-IV graph
requires both abduction and non-stratified negation, as can be seen by
considering the diagnosis of Adjustment Disorder (\cite{DSM-IV},
pg. 626).  One criterion for this diagnosis is
\begin{center}
{\em Once the stressor (or its consequences) has terminated, the
symptoms do not persist for more than an additional 6 months.}
\end{center}
Thus, to diagnose a patient as presently undergoing adjustment
disorder, a physician must hypothesize about events in the future ---
a step naturally modeled with abduction.  Adjustment disorder requires
an exclusion criterion
\begin{center}
{\em The stress-related disturbance does not meet the criteria for
another specific Axis I disorder and is not merely an exacerbation of
a preexisting Axis I or Axis II disorder.}
\end{center}
that admits the possibility of a loop through negation between
adjustment disorder and another diagnosis.  This can in fact occur,
for instance with Alzheimer's Dementia (\cite{DSM-IV}, pg. 142-143).
If, as far as a physician can tell, a patient fulfills all criteria
for adjustment disorder besides the above criterion, as well as all
criteria for Alzheimer's (besides the criterion that the disturbance
is not better accounted for by another disorder), the physician will
essentially be faced with the situation:
\begin{center}
{\em The patient has an Adjustment Disorder if he does not have
Alzheimer's Dementia, and has Alzheimer's Dementia of the patient does
not have an Adjustment Disorder.}
\end{center}
%Indeed, such a situation is possible, for instance in the case of an
%elderly patient about whose disturbance a physician does not have full
%knowledge.
%

Use of abduction over DSM-IV must therefore handle non-stratified
programs.  The current user interface of the Diagnostica system ({\tt
http://medicinerules.com}) uses abduction in a simple but clinically
relevant way to allow for hypothetical diagnosis: when there is not
enough information about a patient for a conclusive diagnosis, the
system allows for hypothesizing possible diagnosis on the basis of the
limited information available.

\paragraph*{Model-based Diagnosis}  \wfsmeth{} has also been employed to
detect specification inconsistencies in model-based diagnosis system
for power grid failure \cite{CasP02}. Here abduction is used to attempt
to abduce hypothetical physically possible events that might cause the
diagnosis system to come up with a wrong diagnosis, violating the
specification constraints. It is akin to model verification: one
strains to abduce a model, comprised of abduced physical events, which
attempts to make the diagnostic program inconsistent.  If this cannot
be done, the power grid can be certified to be correct.
% NEW sentence:
The attempt is conducted by trying to abduce hypothetical real world
events which would lead to a proof of \emph{falsum}, the atom reserved for
the purpose of figuring in the heads of integrity constraints having the
form of denials.

In this case, the application concerns a real electrical power grid
network in Portugal, which is being monitored in real time by a
pre-existing model-based logic programming diagnosis system
(SPARSE)\footnote{cf.  {\tt
http://www.cim.isep.ipp.pt/Projecto-SATOREN/}} that receives
time-stamped event report messages about the functioning or
malfunctioning of the grid. The aim of our abductive application was
to certify that a given expert system diagnosis module was provably
correct with respect to foreseen physical events. To wit, the
diagnosis logic program was executed under \wfsmeth{} in order to
establish that no sequence of (abduced) physically coherent events
(i.e. monitoring messages) could be conducive to a diagnosis violating
the (temporal) constraints expected of a sound diagnosis.

This approach proved to be feasible, though it required us to
introduce a constructive negation implementation of \wfsmeth{}, not
yet reported elsewhere, because the abduced message events had to be
time-stamped with temporally constrained conditions with variable
parameters, and often these occurred under default negated literals
(and hence the need for applying constructive negation on those
variables), to the effect that no supervening event took place in some
time related interval. The system, the application, and its use are
described in detail in \cite{CasP02}.

Four steps were involved in this process:
\begin{itemize}
\item Translation of the SPARSE rules into a syntactical form suitable
for abduction.
\item Preprocessing of the translated rules for use by our \wfsmeth{}
implementation.
\item Obtaining abductive event solutions for diagnosis goals.
\item Checking for physical consistency of the abductive solutions.
\end{itemize}

The most difficult and critical step was the first one, as the
pre-existing SPARSE expert system rules had been written beforehand by
their developers, with no abductive use in mind at all. Specific tools
were developed to automate this step. The pre-existing
\wfsmeth{} implementation (comprising constructive negation) mentioned at
step three (which required minimal adaptation), and the dualization
preprocessor, mentioned at step two, both functioned to perfection.
Step four was enacted by constructing tools to automate the analysis
of the physical meaningfulness of the abduced solutions.

A number of open problems worthy of exploration remain in this class of
problems, susceptible of furthering the use of the general abductive
techniques employed.

\paragraph*{Reasoning about Actions}  \wfsmeth{} has been applied as well
to model and reason about actions \cite{ALPQ00}. For this the
\wfsmeth{} system was integrated with Dynamic Logic Programming
(DLP) Updates system \cite{ALPPP00}.

DLP considers sequences of logic programs $P_1 \oplus P_2 \oplus
\dots P_n$, whose intended meaning is the result of updating the
logic program $P_1$ with the rules in $P_2$, then updating the
resulting knowledge base with \dots, and then updating the
resulting knowledge base with $P_n$. In \cite{ALPPP00} a
declarative semantics for DLP is presented. In order to ease the
implementation of DLP, \cite{ALPPP00} also presents an
alternative, equivalent, semantics which relies on a
transformation of such sequences of programs into a single logic
program in a meta-language. This transformation readily provides
an implementation of DLP (obtainable via {\tt
http://centria.di.fct.unl.pt/\symbol{126}jja/updates}). For this
implementation the use of tabling is of importance. In fact, the
transformation relies on the existence of inertia rules for
literals in the language, stating that some literal is true at
some state if it was true before and is not overridden at that
present state. Tabling is important for the efficiency of the
implementation by avoiding repetition of computation for past
states.

DLP has been used in applications for reasoning about actions
\cite{ALPQ00}. In this setting actions are coded as logic programs
updates which may have pre-conditions and post-conditions. For
these applications the possibility of having programs with loops
over negation is crucial. In fact, rules involved in such loops
are used to model for instance unknown initial conditions and
unknown outcomes of actions. For a concrete example, if one wants
to state that initially it is not known whether or not the
individual $a$ was alive, one may write, in the first program
$P_1$, the rules:

\[\begin{array}{rcl}
alive(a) & \leftarrow & not \neg alive(a)\\
\neg alive(a) & \leftarrow & not\ \!  alive(a)\\
\end{array}\]

Reasoning about actions in a scenario is performed by a
well-founded evaluation of the sequence of updated programs.
Abductive reasoning is used for planning in the actions scenario.
In fact, in this update setting, abducing updates (which code
actions) in order to fulfill some goal of some future state
amounts to plan which actions need to be execute in order to make
that goal true. For this, a system with tabling, ability to deal
with programs with loops over negation, and abduction was needed.
\wfsmeth{} includes all these ingredients, and was successfully
employed for this purpose.

\mycomment{
\paragraph*{Reasoning about Actions}  \wfsmeth{} has been applied as well
to model and reason about actions \cite{ALPQ00}. For this the
\wfsmeth{} system was integrated with Dynamic Logic Programming
Updates \cite{ALPPP00} system, so that updates can be abduced. In
this application (obtainable via {\tt
http://www-ssdi.di.fct.unl.pt/\symbol{126}jja/updates}) actions,
which may have preconditions and post-conditions, are represented
as updates to knowledge bases represented by means of logic
programs. A given scenario is represented by a sequence of
updates, and reasoning about such a scenario is performed by a
well-founded evaluation of the sequence of updated programs.
Abduction of actions (coded as updates) is used for planning.
}

\subsection{Comparisons with Other Methods}

The use of dual programs to compute the well-founded semantics of
normal programs was introduced in \cite{PeAA91}, but this method
has several limitations compared with \wfsmeth{}: it does not
handle abduction or explicit negation; and it can have exponential
complexity for some queries.  Many of the definitions of
\wfsmeth{} are derived from SLG \cite{CheW96} (as reformulated in
\cite{Swif99b}) which computes queries to normal programs
according to the well-founded semantics.  For normal programs,
\wfsmeth{} shares the same finite termination and polynomial
complexity properties as SLG. \wfsmeth{} adds the capability to
handle abduction (by adding abductive contexts to goals, modifying
operations on forests to deal with such contexts, and by adding
the {\sc Abduction} operation), adds the use of the dual
transformation for extended programs and the \unfr{} operation,
but \wfsmeth{} does not allow evaluation of a non-ground program
as does SLG. Unfortunately, performance trade-offs of \wfsmeth{}
and SLG are not yet available, due to the lack of an engine-level
implementation of the \unfr{} operation of \wfsmeth.

The main contribution of \wfsmeth{} is its incorporation of
abduction. We are not aware of any other efforts that have added
abduction to a tabling method.  Indeed, it is the use of tabling
that is responsible for the termination and complexity results of
Sections \ref{sec:term-compl} and \ref{sec:gsm}. Furthermore,
\wfsmeth{} evaluations are confluent in the sense that Theorem
\ref{thm:exdual} holds for {\em any} ordering of applicable
\wfsmeth{} operations.  The complexity and termination for $WFS$
distinguishes \wfsmeth{} from approaches such as the IFF proof
procedure \cite{FunK96} and SLDNFA \cite{DeDe98}.  At the same time,
these approaches do allow variables in rules which \wfsmeth{} does
not.  The methods of \cite{cdt91} and \cite{Insa99} compute abductive
explanation based on some form of two-valued rule completion for
non-abducible predicates (the former based on Clark's completion, and
the latter based on the so-called transaction programs). This is
similar to our use of the dual program\footnote{Note that the dual for
non-abducible predicates in acyclic programs is the same as the
completion.}. In both methods, abductive explanations are computed by
using the only-if part of the completion in a bottom-up
fashion. However, both methods have a severe restriction on the class
of programs: they apply generally only to acyclic programs.  This
restriction is due to their being based on completion so that from an
\wfsmeth{} perspective, these methods do not require operational
analogs to the \delay{} and \simpl{} operations to evaluate unfounded
sets of objective literals, or the \unfr{} operation to evaluate
co-unfounded sets.  The pay-off of adding these operations is that
\wfsmeth{} is based on the well-founded semantics, and does not impose
any restriction on cycles in programs.

The restriction on cycles is also not imposed by methods based on
the stable models semantics, such as \cite{Sa91,Sa00,KaMo97}. As
\wfsmeth{}, the method of \cite{Sa91} also requires a prior
program transformation. In this case, an abductive programs is
translated into a normal logic program, such that the stable
models of the latter correspond to the abductive solutions of the
former. This method has some drawbacks. Most importantly, by doing
so, one may obtain abductive solutions with atoms that are not
relevant for the abductive query. To avoid this drawback, in
\cite{Sa00} the method is improved by incorporating a top-down
procedure to determine the relevance of the abducible to the
query. The ACLP system of \cite{KaMo97} is based on Generalized
Stable Models, but it also integrates in a single framework
abduction and constraint programming. Again the complexity results
for $WFS$, when compared to that for the stable models semantics,
distinguishes \wfsmeth{} from these approaches.

We have shown in this article how \wfsmeth{} can be mustered to
compute Generalized Stable Models, and thus Stable Models in
particular. Some words are in order on comparing it to other
Stable Model implementations, such as DLV \cite{dlv}, and S-Models
\cite{smodels}. These implementations are specialized toward
Stable Model evaluation, and are restricted to finite ground
programs without functional symbols, though some preprocessors can
help to do the grounding where possible and domain information is
available. Naturally, their efficiency for the specific purpose of
computing Stable Models is better than that of a general procedure
like \wfsmeth{}, even though the complexity remains the same.

Abduction can also be carried out by those specialized
implementations by means of known program transformations, such as
the ones shown in \cite{LT:NMELP95,SI:JLP00}. Though one common
problem to those approaches is that, because of the non-relevancy
character of Stable Models, and also of abducibles being
two-valued in them, all possible (non-minimal) abductions are
potentially generated, and not just those relevant for a top goal.

With respect to stratified programs, where the well-founded and
stable models semantics coincide, \wfsmeth{} is able to deal with
function symbols and non-ground programs in infinite domains, and
perform demand driven abduction. Moreover, if abduction is not
after all required, then the complexity, we have seen, remains
polynomial, and no unnecessary abductions are made, in
contradistinction to the two-valued approach, which requires for
all abducibles to be abduced either as true or as false.

In summary, the two approaches are designed for different purposes,
and each should excel in its own territory.  Proctracted attempts to
have \wfsmeth{} compute a relevant residual program that would be passed
on to an implementation of stable models have failed, as most of the
work ends up having to be done on the \wfsmeth{} side, without the desired
sharing of specialized effort.

\paragraph*{Generalizing \wfsmeth{} to Programs with Variables}
Generalizing \wfsmeth{} for non-ground covered programs\footnote{A
program is covered iff all variables appearing in the body of rules
also appear in the corresponding head.} with ground queries is not a
difficult task: as in Clark's completion, consider rule heads with
free variables, and explicitly represent unifications in the body; the
dual is then obtained from these rules as usual, where the negation of
\verb+=+ is \verb+\=+. Allowing non-ground queries in covered programs
can be obtained by considering as abducibles all terms of the form
\verb+X \= T+, and by adding an appropriate method for verifying
consistency of sets of such inequalities.
%Note that this allows for constructive negation (in covered programs).
Such a method could greatly benefit from an integration of
\wfsmeth{} with constraint programming, where the consistency of
the inequalities would be checked by a constraint solver. The
integration of abduction and constraint programming, as is already
done for other systems (viz. SLDNFAC \cite{NuDe00} and ACLP
\cite{KaMo97}), is in our research agenda.

The most difficult step in order to fully generalize \wfsmeth{} to
deal with non-ground programs is to abandon the restriction of covered
programs.  This is so because free variables in the body of program
rules introduce universally quantified variables in the body of rules
in the dual program --- a problem similar to that of floundering in
normal programs.  Work is underway to generalize \wfsmeth{} to deal
with non-ground non-covered programs using constructive negation
methods.

A practical advantage of \wfsmeth{} is that it allows the easy
propagation of abducibles through both positive and negative literals.
As an abductive answer is returned to an abductive subgoal, contexts
can be immediately checked for consistency, regardless of whether the
subgoal is positive or negative, and regardless of how many levels of
negation were needed to produce the answer.

{\large {\bf Acknowledgements}} This work was partially supported by
NSF grants CCR-9702581, EIA-97-5998, and INT-96-00598.  The authors
also thank PRAXIS XXI projects FLUX and FLAD-NSF project REAP for
their support.  Finally, the presentation of this paper was
considerably improved with the help of detailed comments from an
anonymous reviewer.

\appendix
\section{Appendix: Proofs of Theorems}

\subsection{Proof of Theorems in Section~\ref{sec:snc} }

\noindent
{\bf Theorem \ref{thm:dual-equiv}} {\em For a ground program $P$ and
empty set of abducibles, let $dual(P,\emptyset)$ be the dual program
formed by applying the dual transformation to $P$.  Then for any
literal $L \in literals(P)$,
\[
L \in WFS(P) \iff L \in lfp(\omega_d^{dual(P,\emptyset)})
\]
}
\begin{proof}

As mentioned in Section \ref{sec:wfs}, if $P$ is a countable set of
rules, each with a finite number of literals in their bodies,
$dual(P,\emptyset)$ will be also.

The inner fixed point of $\omega_d^{dual(P,\emptyset)}$ depends on two
operators: $Td_{\cI}^{dual(P,\emptyset)}$ and $Fd_{\cI}^{dual(P,\emptyset)}$
(Definition~\ref{def:dualintops}).  $Td_{\cI}^{dual(P,\emptyset)}$ is
essentially the same as the inner fixed point operator, $Tx^P_{\cI}$
(Definition~\ref{def:intops}) used to construct WFS(P).  Combining
this observation with the initial sets used to construct the fixed
points of Definitions~\ref{def:fp_inner} and \ref{def:dualfp_inner},
the proof can be reduced to showing that for an element $O$ in
$objective\_literals(P)$,
\[
O \in gfp(Fx^P_{\cI}(objective\_literals(P))) \iff not(O) \in
 	gfp(Fd^{dual(P,\emptyset)}_{\cI}(\cS_{dual(P,\emptyset)})) 
\]
If this is so, since $Tx^P_I$ equals $Td^{dual(P,\emptyset)}_{\cI}$
restricted to the objective literals in $P$, then it is a trivial
induction on the operators of Definitions~\ref{def:fp_inner} and
\ref{def:dualfp_inner} to show that $\omega^P_{ext}$ equals
$\omega^{dual(P,\emptyset)}_{d}$ restricted to the literals in $P$.

Note that the initial set used to construct the fixed point of
$Fx^P_{\cI}$ in Definition~\ref{def:fp_inner} is
$objective\_literals(P)$, while in Definition~\ref{def:dualfp_inner}
the initial set used for $Fd_{\cI}^{dual(P,\emptyset)}$ is
\[
\cS_{dual(P,\cA)} = \{not(O) | not(O) \in literals(dual(P,\cA)) \} 
\]
and thus by the dual transformation an objective literal $O$
occurring in $P$, $O$ is in $objective\_literals(P)$ iff $not(O)$ is
in $\cS_{dual(P,\cA)}$.

($\Rightarrow$) We first prove that for $O \in objective\_literals(P)$:
\[
O \in gfp(Fx^P_{\cI}(objective\_literals(P))) \Rightarrow not(O) \in
 	gfp(Fd^{dual(P,\emptyset)}_{\cI}(\cS_{dual(P,\emptyset)}))
\]
\mycomment{
The proof of this statement is complicated by the fact that
$Fd^{dual(P,\emptyset)}$ need not be continuous, so that
$gfp(Fd^{dual(P,\emptyset)}_{\cI}(\cS_{dual(P,\emptyset)}))$ may arise
only after a transfinite number of steps.  
}

Induction is on the number $n$ of applications of $Fx^P_{\cI}$ in
constructing the fixed point. The base case, where $n = 0$ was handled
in the previous paragraph, so consider first the case in which $n$ is
greater than $0$.  In other words, the two operators have been applied
$n-1$ times, with applications of $Fx^P_{\cI}$ producing the set
$\cO_1$ and applications of $Fd^{dual(P,\emptyset)}_{\cI}$ producing
the set $\cO_2$.  It remains to prove that for the $n^{th}$
application:
\[
O \in Fx^P_{\cI}(\cO_1)) \Rightarrow not(O) \in 
 	Fd^{dual(P,\emptyset)}_{\cI}(\cO_2)
\]
Suppose an objective literal $O$ is in $Fx^P_{\cI}(\cO_1)$.  Then
either (1) $conj_E(O) \in \cI$; or for every rule $r^O_j$ of the form
$O \mif{} L_1,...,L_n$ in $P$, there is a literal $L_{j,i}, 1 \leq i
\leq n$, such that either (2a) $conj_D(L_{j,i}) \in \cI$ or (2b)
$L_{j,i}$ is in $\cO_1$.  Consider each of these cases in turn.  
\begin{enumerate}
\item For the first case, if $conj_E(O) \in \cI$ then, by the rule $not(O)
\mif conj_E(O)$ in $dual(P,\emptyset)$, $not(O)$ belongs to
$Fd^{dual(P,\emptyset)}_{\cI}(\cO_2)$.  
\item For the second case, consider
a witness of unusability, $L_{j,i}$ for a rule $r^O_j$ for $O$.  (2a)
Suppose first that $conj_D(L_{j,i}) \in \cI$.  Then there is a folding
rule $not(fold^b_j\_O)\mif{} conj_D(L_{j,i})$ constructed by the dual
transformation of Definition~\ref{dual-fold}, so that
$not(fold^b_j\_O)$ will be included in
$Fd^{dual(P,\emptyset)}_{\cI}(\cO_2)$.  (2b) Alternately, if $L_{j,i}
\in \cO_1$, then $not(L_{j,i}) \in \cO_2$ so that $not(L_{j,i})$ and
all literals in the heads of the folding rules that depend on it will
be regenerated.  In either case (2a) or (2b), by monotonicity of the
operators, if $O \in Fx^P_{\cI}(\cO_1))$, then $O \in \cO_1$, and by
the induction hypothesis, $not(O) \in \cO_2$.  By definition of
$Fd^{dual(P,\emptyset)}_{\cI}$, for this to happen each rule for $O$
in $P$ must have a witness of unusability.  This means that the
literal $not(fold^a\_O_k)$ must also be in $\cO_2$ for each rule $k$
for $O$ in $P$.  Thus each of the literals $not(fold^b\_O_l)$ and
$not(fold^b\_O_k)$ will be regenerated, so that $not(O) \in
Fd^{dual(P,\emptyset)}_{\cI'}(\cO_2)$.
\end{enumerate}

($\Leftarrow$) To complete the proof we need to show that for $O \in
objective\_literals(P)$:
\[
O \in gfp(Fx^P_{\cI}(objective\_literals(P))) \Leftarrow not(O) \in
 	gfp(Fd^{dual(P,\emptyset)}_{\cI}(\cS_{dual(P,\emptyset)}))
\]

The proof of this statement is complicated by the fact that, due to
folding rules, there is no exact correspondance between the iteration
in which an objective literal is removed from a set by $Fx^P_{\cI}$
and when its analog is rmoved by $Fd^{dual(P,\emptyset)}_{\cI}$.
Accordingly, we restate this case as
\[
O \not\in gfp(Fx^P_{\cI}(objective\_literals(P))) \Rightarrow not(O) \not\in
gfp(Fd^{dual(P,\emptyset)}_{\cI}(\cS_{dual(P,\emptyset)})) 
\]
Consider a literal $O'$ in 
\[
objective\_literals(P) - gfp(Fx^P_{\cI}(objective\_literals(P)))
\]
Because $Fx^P_{\cI}$ is monotonic and continuous, there is an
application $n$ such that $O' \in
gfp(Fx^{P^{n-1}}_{\cI}(objective\_literals(P)))$ and $O' \not\in
gfp(Fx^{P^n}_{\cI}(objective\_literals(P)))$.  We call $n$ the {\em
removal level} of $O$.

Thus we prove by induction on $n$ that any objective literal with
removal level $n$ is not in
$gfp(Fd^{dual(P,\emptyset)}_{\cI}(\cS_{dual(P,\emptyset)}))$.  The
base case (where $n = 0$) that for $O \in objective\_literals(P)$, $O
\in objective\_literals(P) \iff not(O) \in
\cS_{dual(P,\emptyset)}$ was handled above.  Assume that the property
holds for objective literals in $P$ whose removal level is less than
$n$, and that the set produced by the first $n-1$ applications of
$Fx^P_{\cI}$ is $\cO_1$.

Consider a literal $O$ with removal level $n$.  By
Definition~\ref{def:intops}, $O$ will not persist after the $n$th
application of $Fx^P_{\cI}$ iff $conj_E(O) \not\in I$ and there exists
a rule $r^k_O$ with body $L_1,...,L_m$ such that for all $1
\leq i \leq m$, $conj_D(L_i) \not\in \cI$ and $L_i \not\in \cO_1$.  

First note that $conj_E(O) \not\in I$ means that an axiom of coherence
will not be used to derive $not(O)$ in
$gfp(Fd^{dual(P,\emptyset)}_{\cI}(\cS_{dual(P,\emptyset)}))$.  Next,
by Definition~\ref{dual-fold}, there is a folding rule
\[
not(fold^b_k\_O) \mif conj_D(L_i)
\]
for each $L_i$ in the body of $r^k_O$.  Note that $conj_D(L_i)$ is not
in $\cI$ (otherwise $O$ would not have a removal level).  If $L_i
\not\in \cO_1$, then the removal level of $L_i$ is less than $n$ so by the
induction assumption, there will be some application of
$Fd^{dual(P,\emptyset)}_{\cI}$ that does not regenerate $conj_D(L_i)$.
Accordingly, there will be an application of
$Fd^{dual(P,\emptyset)}_{\cI}$ in which $not(fold^b_k\_O)$ is not
regenerated via its $i$th clause.  For $O$ to have a removal level,
each literal in the body of $r^k_O$ must have similar conditions to
$L_i$, and $not(fold^b_k\_O)$ is not regenerated.  By
Definition~\ref{dual-fold}, regeneration of $not(fold^b_k\_O)$ is
required to regenerate $not(fold^a_k\_O)$.  In future iterations,
then, $not(fold^a_j\_O)$ will not be regenerated for $1 \leq j \leq
k$, so that $O$ will not be in
$gfp(Fd^{dual(P,\emptyset)}_{\cI}(\cS_{dual(P,\emptyset)}))$.

%------------------

\end{proof}

%---------------------------------------------------------------------

In addition to Theorem~\ref{thm:dual-equiv}, several lemmas and
definitions will be needed to prove the correctness of \wfsmeth{}.
One of these is a simpler definition of the dual transformation, which
is convenient to use in the proofs.  This definition was implicitly
used in Examples~\ref{ex1a},~\ref{ex1} and~\ref{ex2}.

\begin{definition} [{\bf Unfolded dual Program}] \label{dual-unfold}
Let $P$ be a ground extended program, and $\cA$ a (possibly empty) set
of finite abducibles.  Then the {\em unfolded dual transformation}
creates a dual program $udual(P, \cA)$, defined as union of $P$ with
the smallest program containing the sets of rules $R_1$ and $R_2$
defined as follows:
\begin{enumerate}
\item If $P$ contains a rule with non-empty body 
\[
\begin{array}{rl}
	O\mif & L_{1,1},...,L_{1,n_1}\\ : & \\ 
	O\mif & L_{m,1},...,L_{m,n_m}
\end{array}
\]
Then, $R_1$ contains the rules 
\[
\begin{array}{rl}
	not(O) \mif & conj_D(L_{1,j_1}), ..., conj_D(L_{m,j_m}).
\end{array}
\]
for each $j$, $1 \leq j_i \leq i_m$, $1 \leq i \leq m$, and where
$conj_D(L)$ represents the default conjugate of $L$.

\item Otherwise, if $not(O)$ is in $literals(P)$, but there is no rule
with head $O$ in $P$, then $R_1$ contains the rule $not(O) \mif{} {\bf t}$. 
\item $R_2$ consists of {\em axioms of coherence} that relate explicit
and default negation, defined as:
\[
not(O) \mif conj_E(O)
\]
For each objective literal $not(O)$ in either $literals(P \cup R_1)$
or $\cA$.
\end{enumerate}
\end{definition}

%-------------------------------------------------------
\noindent
Because the dual transformation of Definition \ref{dual-fold} differs
from that of Definition \ref{dual-unfold} only insofar as no folding
rules are defined, it is straightforward to see that they are
equivalent with respect to $literals(P)$ for finite ground programs.
While Definition~\ref{dual-unfold} is simpler than
Definition~\ref{dual-fold} and useful for correctness results,
Definition~\ref{dual-fold} is necessary for proving correctness of
programs with an infinite number of rules and for the complexity
results that follow.

%-------------------------------------------------------

We next make explicit the relation between unfounded sets of objective
literals in $P$ and co-unfounded sets of answers in the dual program.
In order to do so, we present the definition of an interpretation
induced by the state of an \wfsmeth{} evaluation, and of the delay
dependency graph.

%-------------------------------------------------------
\begin{definition} \label{def:induced-interp}
Given an \wfsmeth{} forest $\cF$, the {\em interpretation induced by
$\cF$}, or $\cI_{induced(\cF)}$ is defined as the smallest
interpretation containing
\begin{itemize}
\item A literal $O$ for each unconditional answer node $<O,\emptyset > \mif
|$ in $\cF$; 
\item A literal $not(O)$ for each objective literal $O \in \cF$ such that the
tree for $O$ is completely evaluated in $\cF$ and contains no answers.
\end{itemize}
\end{definition}
%-------------------------------------------------------

The definition of a delay dependency graph is convenient for several
of the following proofs.

\begin{definition} [{\bf Delay Dependency Graph}] \label{def:ddg} 
Let $\cF$ be an \wfsmeth{} forest.  A goal $S_1$ has a {\em direct
delay dependency on} a goal $S_2$ in $\cF$ iff $S_2$ is contained in
the $DelayList$ of an answer in the tree $T$ for $S_1$.  The
\emph{delay dependency graph} of $\cF$, \emph{DDG($\cF$)}, is a
directed graph \emph{\{V,E\}} in which $V$ is the set of root goals
for trees in $\cF$ and $(S_i,S_j) \in E$ iff $S_i$ has a direct delay
dependency on $S_j$.
\end{definition}

Also for convenience, if $\cS$ is a co-unfounded set of answers, 
\[
heads(\cS) = \{S| <S,Context> \mif DL| \in \cS \}
\]

Part (1) of Lemma~\ref{lem:co-unfounded-wfs} relates a co-unfounded
set of answers (Definition~\ref{def:co-unfounded}) obtained in the
\wfsmeth{} evaluation of a ground extended program to an unfounded set
of objective literals (Definition~\ref{def:unfounded}) in the
well-founded semantics (with explicit negation).  Intuitively, part
(2) ensures that when an \wfsmeth{} forest can be constructed to
capture an interpretation $\cI$ of a ground extended program, then the
forest can be extended to determine the truth value of each objective
literal in the unfounded sets with respect to $\cI$.

%-------------------------------------------------------
\begin{lemma} \label{lem:co-unfounded-wfs}
Let $< P,\cA,I >$ be an abductive framework in which $\cA$
and $I$ are empty.
\begin{enumerate}
\item   Let
$\cF$ be a forest in a \wfsmeth{} evaluation $\cE$ of a query $Q$ to
$< P,\cA,I >$, and $S^A$ be a co-unfounded set of answers
in $\cF$.  Then there is a minimal unfounded set of objective literals
$S^O$ for $P$ in $\cI_{induced(\cF)}$ such that $\{S_i | not(S_i)
\in heads(S^A)\} \subseteq S^O$.
\item Let $S^O$ be a minimal unfounded set of objective literals for
$P$ with respect to an
% consistent 
interpretation $\cI$ and let $\cF$ be an \wfsmeth{} forest of any
query to $< P,\cA,I >$ such that
\begin{enumerate}
\item for each $S_i \in S^O$, there is a tree for $not(S_i)$ in $\cF$;
\item for all  $L \in I$, if $<L,\emptyset>$ is the root of a tree in
$\cF$, then $L \in I_{induced(\cF)}$;
\item no \wfsmeth{} operations are applicable to $\cF$.
\end{enumerate}
Then, for each $S_i \in S^O$, there will be an unconditional answer for
each $not(S_i)$ in $\cF$.
\end{enumerate}

\end{lemma}
\begin{proof}
For simplicity of presentation, we first restrict our attention to
finite programs, in which class the dual program,
$udual(P,\emptyset)$, formed via Definition~\ref{dual-unfold} can be
used.  We then indicate how the arguments can be extended to the dual
form of Definition~\ref{dual-unfold} which is necessary for infinite
programs.
\begin{enumerate}
\item Let $N_{leaf} = <not(S_i),\emptyset> \mif DL |$ be an answer in
some co-unfounded set of answers $S^A$ in $\cF$.  By clause 3 of
Definition~\ref{def:co-unfounded}, $DL$ must be non-empty.  By the
construction of Definition~\ref{dual-unfold} and by the definitions of
\wfsmeth{} it can easily be seen that the node $<not(S_i),\emptyset>
\mif DL |$ is a descendant of a non-root node $N =
<not(S_i),\emptyset> \mif L_1,...,L_n$ produced by application of a
\pgmcr{} operation of a rule $r_u$ in $udual(P,\emptyset)$ (this situation is
presented schematically in  Figure~\ref{fig:schema})  $r_u$ is
constructed so that for each rule $r_j$ with head $S_i$ in $P$, $L_j$
is a default conjugate of some literal in the body of $r_j$.  Now we
consider two classes of the literals $L_1,...,L_n$ in the ($GoalList$
of the) node $N$: those that are contained in the $DelayList$ of
$N_{leaf}$ and those that are not, and consider the latter first.

%------------------------------------------------------------------
\begin{figure}[hbtp] 
\centering
	\fbox{\epsfig{file=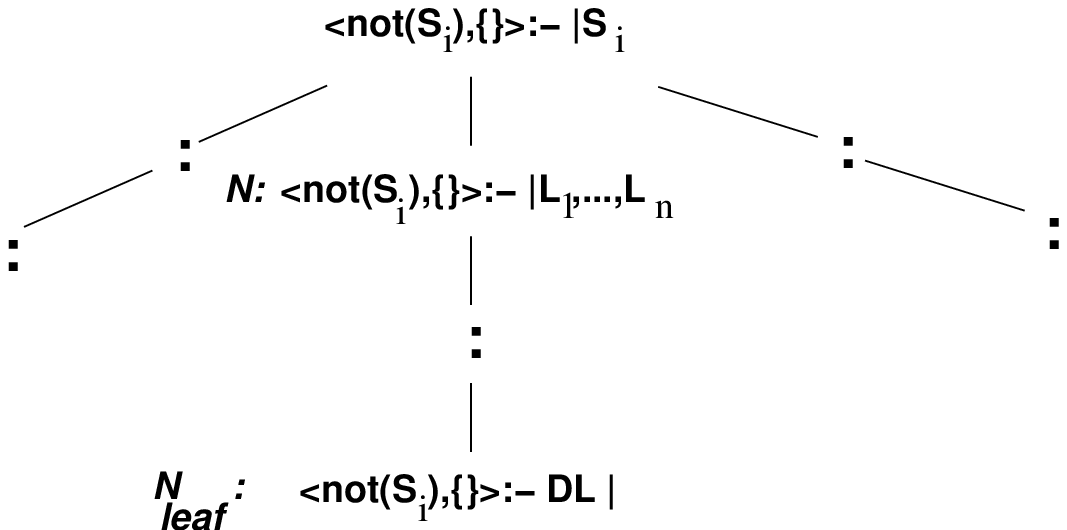,width=0.65\textwidth}}
\caption{Schematic portion of an \wfsmeth{} tree}
\label{fig:schema}
\end{figure}
%------------------------------------------------------------------

\begin{enumerate}
\item Literals that are in $N$ but not in the $DelayList$ of $N_{leaf}$.  
%By condition (1) of Definition \ref{def:co-unfounded}, $not(S_i)$
%is completely evaluated and 
By assumption, since the set of abducibles $\cA$ is empty, there can be
no {\sc Abduction} operations in $\cE$; thus a literal $L_i$ in $N$
but not in $N_{leaf}$ can have been resolved away either through
\anscr, or through \delay{} and subsequent \simpl{}.  Note that $L_i$
cannot be created through a direct application of \unfr{} as that
operation acts only on leaf nodes, and removes all elements in a
$DelayList$, contradicting the fact that $DL$ is empty.  In other
words, $L_i$ could be resolved away only via an \anscr{} or \simpl{}
operation, and these operations would be possible only if $L_i$ were
in the interpretation induced by a previous forest, say $\cF_i$.
Furthermore, since no \wfsmeth{} operation in Definition
\ref{def:ops} allows operations on an unconditional answer node, all
answers in $\cF_i$ will persist in $\cF$, so that $L_i$ must be in
$\cI_{induced(\cF)}$.  It is then straightforward from
Definitions~\ref{dual-unfold} and~\ref{def:unfounded} that if $L_i \in
I_{induced(\cF)}$, then its conjugate will form a witness of
unusability for some rule for $S_i$ in $P$.
\item Literals that are in $N$ and in the $DelayList$ of
$N_{leaf}$.  Because $N_{leaf} \in S^A$, each literal, $L_i \in DL$ is
such that $<L_i,C_i> \mif DL_i| \in S^A$, by
Definition~\ref{def:co-unfounded}.
\end{enumerate}

Taken together, the above two cases show that, given an conditional
answer $ <not(S_i),\emptyset> \mif DL \in S^A$, a witness of unsuitability
can be obtained for every rule $r_i$ for $S_i$ in $P$.  This can occur
because $r_i$ contains some literal $conj_D(L)$ that is true in
$\cI_{induced(\cF)}$, corresponding to condition (1) of
Definition~\ref{def:unfounded}; or it can occur because the literal
belongs to $heads(S^A)$, corresponding to condition (2) of
Definition~\ref{def:unfounded}.  Thus, in $S^A$, a witness of
unusability can be obtained for every rule for $S_i$, so that $\{S_i|
not(S_i) \in heads(S^A) \}$ is an unfounded set for
$\cI_{induced(\cF)}$ in $P$.
\item 
Since $S^O$ is a minimal unfounded set of objective literals, no
objective literal $O_1$ in $\cI$ can have its default conjugate in
$S^O$ (otherwise, $S^O - \{O_1\}$ would also be minimal).  Let $S_i$
be an objective literal in $S^O$, and let $L_i, 1 \leq i
\leq n$, be witnesses of unusability for each of its ($n$) rules.  By
the construction of Definition~\ref{dual-unfold}, there is a rule
$r_{S_i} = not(S_i) \mif conj_D(L_1),...,conj_D(L_n)$ in
$udual(P,\emptyset)$.  By assumption (a), $\cF$ contains a tree for
$not(S_i)$.  By assumption (c) and Definition~\ref{def:ops}, this tree
must have children, otherwise there would be \pgmcr{} operations
applicable for $\cF$.
%TLS: since not(S_i) is in a co-unf set, it must have at least one rule.
In particular, there must be a node $N' = <not(S_i),\emptyset> \mif
conj_D(L_1),...,conj_D(L_n)$, that is a child of the root node for
$not(S_i)$.  

Now consider each $conj_D(L_i)$.  Since $L_i$ is a witness of
unusability for $S^O$, $L_i$ is either be false in $\cI$ or unfounded
but not false in $\cI$.
\begin{enumerate}
\item If $L_i$ is false in $\cI$, then by assumption (b) $conj_D(L_i)$
is true in $\cI_{induced(\cF)}$, and either an \anscr{} or \simpl{}
(after previous \delay ) will be applicable, by assumption (c)
these will have been performed and there will be a descendant of $N'$
in which $conj_D(L_i)$ is resolved away.
\item Next, consider the case in which $L_i$ is unfounded but not
false in $\cI$, and by assumption is in $S^O$.  Because $L_i$ is in an
unfounded set it is a positive literal, and $conj_D(L_i)$ can be
written as $not(L_i)$.  Consider the node $N'$ mentioned above:
$<not(S_i),\emptyset> \mif conj_D(L_1),...,conj_D(L_n)$ for which
$not(L_i)$ is a body literal.  For $1 \leq i \leq n$, each
$conj_D(L_i)$ is either in $\cI_{induced(\cF)}$ (by assumption (b))
and resolved away by (assumption (c)); or its default conjugate is
unfounded in $\cI$ and thus a negative literal, so that there is a
node $not(S_i) \mif DL$ in a previous forest $\cF_{prev}$ of $\cF$ in
$\cE$, and $not(L_i) \in DL$ (by assumption (b)).

We must show that a \unfr{} operation was applicable that removed
$not(L_i)$ from $DL$.  For this to happen, we must show that
$not(L_i)$ was part of a co-unfounded set of answers for some previous
forest $\cF_{prev}$ in $\cE$.  Condition (2) of the definition of a
co-unfounded set of answers (Definition~\ref{def:co-unfounded}) is
trivially satisfied since the set of abducibles is empty, while
condition (1) of Definition~\ref{def:co-unfounded}, stating that goals
in the co-unfounded set of answers be completely evaluated, is
satisfied by assumption (c).  Condition (3) of
Definition~\ref{def:co-unfounded} remains, and we must show that there
is a co-unfounded set of answers in $\cF_{prev}$ containing an answer,
$N_{L_i} = not(L_i) \mif{} DL_i|$ for $not(L_i)$.  We begin by showing
that $N_{L_i}$ exists.  Assumption (c) ensures that a tree for
$not(L_i)$ exists in $\cF$ (by the argument above, $not(L_i)$ was
selected, and assumption (c) ensures that a \newsg{} is performed when
$not(L_i)$ was selected).  Furthermore, assumption (c) implies that
there are no applicable \wfsmeth{} operations for this tree.
Furthermore, $DL_1$ contains all literals of a rule in the unfolded
dual of $P$ that gave rise to $N_L$, but which are not themselves in
$\cI$.
%By Definition~\ref{dual-unfold} there must be at least one node of the
%form $<not(L_i),\emptyset> \mif DL|$ such that every member of $DL$ is the
%default conjugate of a member of $S'$.  
Extending this argument for all elements in the transitive closure of
$not(L_i)$ in the delay dependency graph of $N$
(Definition~\ref{def:ddg}), shows that $L_i$ is contained in an
unfounded set of objective literals.  Thus a \unfr{} operation was
applicable to $\cF_{prev}$ which made the answer for some $not(S_j)$,
$S_j \in S'$ unconditional, and it can be easily seen that this
operation made further {\sc Simplification} operations applicable
based on the unconditional answer for $not(S_j)$.  Furthermore, by
Definition~\ref{def:ops}
% because $I$ is consistent, 
each {\sc Simplification} operation for literal $L$ and conditional
answer $Ans \in S$ made applicable after the \unfr{} operation will
remain applicable until $L$ is removed from the $DelayList$ of $Ans$.
Because the set of unconditional answers for a forest only grows
monotonically, the statement holds.
\end{enumerate}
%The case in which $S^O$ is a non-minimal unfounded set of of $P$ in
%$\cI$ is a simple extension of the case in which $S^O$ is minimal.
\end{enumerate}

Extending the proofs to the dual of infinite programs means that the
dual transformation of Definition~\ref{dual-fold} must be used.  In
the case of infinite programs, there may be an infinite number of
witnesses of unusability for a given objective literal $O$ and
interpretation $\cI$.  Unlike the situation presented above for finite
programs, the witnesses of unusability may be distributed among
different folding trees.  With this complication the argument for part
1 of the lemma can be straightforwardly extended to the situation
where folding rules are used.  In part 2, note that assumptions 2a and
2b together imply that if $O$ is an objective literal in $P$, then
$\cF$ will also contain the appropriate folding literals for
$not(O)$.  Again, the argument can be straightforwardly extended to
this new situation.
\end{proof}
%{lem:co-unfounded-wfs}

%--------------------------- ----------------------------
\begin{lemma} \label{dual-wfs}
Let $< P,\emptyset, \emptyset >$ be an abductive
framework.  Let $\cE$ be a \wfsmeth{} evaluation of a query $Q$
against $udual(P,\emptyset)$, whose final forest is $\cF_{\beta}$.
Finally, let $WFS(P)|_{\cE_{\beta}}$ denote the well-founded model of
$P$ restricted to goals in $\cF_{\beta}$.  Then
\[
\cI_{induced(\cF_{\beta})} =  WFS(P)|_{\cF_{\beta}}
\]
\end{lemma}
\begin{proof}

(Sketch) Given Theorem~\ref{thm:dual-equiv} this is equivalent to
showing that 
\[
\cI_{induced(\cF_{\beta})} =  lfp(\omega_d^{dual(P,\emptyset)})|_{\cF_{\beta}}.
\]
%Note that the operator $\omega_{d}^{dual(P,\emptyset)}$ is defined in
%terms of the operators $Td^{dual(P,\emptyset)}_{\cI}$ and
%$Fd^{dual(P,\emptyset)}_{\cI}$ (Definition~\ref{def:dualfp_inner}).
Since by assumption abduction is not needed, and since the case of
co-unfounded sets was handled in Lemma~\ref{lem:co-unfounded-wfs},
proving that \wfsmeth{} computes the fixed points specified by these
operators is similar to (transfinite) inductions for soundness and
completeness of other tabled evaluations of the well-founded semantics.
\cite{Swif99b} and other papers contain detailed inductions that show that 
the interpretation induced by the final forest of a tabled evaluation
is equivalent to the model preoduced by the least fixed point of the
operator $\omega^P_{ext}$.  Given Theorem~\ref{thm:dual-equiv} and
Lemma~\ref{lem:co-unfounded-wfs}, extending such a proof to abdual is
tedious but straightforward.
\end{proof}
%-------------------------------------------------------

The next step is to extend the results of
Lemma~\ref{lem:co-unfounded-wfs} to arbitrary abductive frameworks
with non-empty sets of abducibles and integrity rules.  The results
must now prove equivalences to models based on abductive scenarios.
In the definition of abductive scenarios
(Definition~\ref{def:ab-scen}) a program $P_\cB$ is constructed based
on a subset of abducible objective literals.  A subset $\cB$ of the
abducibles of an abductive framework $\sigma$ can be obtained directly
from the context, $Context$, of an abductive subgoal; and a program
$P_{\cB}$ can be generated from $\cB$ as in
Definition~\ref{def:ab-scen}.  In the following lemma, we refer to the
program produced by the construction of Definition~\ref{def:ab-scen}
on the objective literals of $Context$ simply as $P_{Context}$, and to
$Context$ seen as an interpretation as $\cI_{Context}$.

%-------------------------------------------------------
\begin{lemma} \label{lem:co-unfounded-wfsab}
Let $< P,\cA,Int >$ be an abductive framework.
\begin{enumerate}
\item Let $\cF$ a forest in an \wfsmeth{} evaluation of a query $Q$
against $< P,\cA,Int >$; $S^A$ be a co-unfounded set of
answers in $\cF$; and let $Context =
\bigcup\{C_i | <S_i,C_i> \mif DL| \in S^A\}$.  Then there is a minimal
unfounded set $S^O$ for $P \cup P_{Context} \cup Int$ with respect to
$(\cI_{induced(\cF)} \cup \cI_{Context})$ such that $\{S_i | not(S_i)
\in heads(S^A)\} \subseteq S^O$.
\item Let $< P,\cA,\cB,Int >$ be an abductive scenario,
and $\cI$ an interpretation of $P \cup P_{\cB} \cup Int$, such that
$\cI|_{\cA}$ is consistent.  Let $S^O$ be an unfounded set of
objective literals with respect to $\cI$.  Let $\cF$ be a \wfsmeth{}
forest of any query to $< P,\cA,Int >$ such that
\begin{enumerate}
\item for each $S_i \in S^O$, there is a tree
for $not(S_i)$ in $\cF$;
\item for all  $L \in \cI$, if $<L,\emptyset>$ is the root of a tree in
$\cF$, then $L \in \cI_{induced(\cF)}$; and 
\item no \wfsmeth{} operations are applicable to $\cF$.
\end{enumerate}
Then $\cF$ will contain an unconditional answer
\[
<not(S_i),C_i> \mif |
\]
for each $S_i \in S^O$, such that $P_{\cup_{i \in S_i} C_i} \subseteq
P_{\cB}$. 
\end{enumerate}
\end{lemma}
\begin{proof}
As in Lemma~\ref{lem:co-unfounded-wfs} we first restrict our attention
to finite programs, for all of which the dual program, $udual(P,\cA)$,
formed via Definition~\ref{dual-unfold} can be used.  We then indicate
how the arguments can be extended to the dual form of
Definition~\ref{dual-unfold} which is necessary for infinite programs.
\begin{enumerate}
\item Let $N_{leaf} = <not(S_i),C_i> \mif DL |$ be an answer in
the co-unfounded set of answers $S^A$.  By the construction of
unfolded dual transformation (Definition~\ref{dual-unfold}), and by
the definitions of \wfsmeth{} it can also be seen that the node
$<not(S_i),C_i> \mif DL |$ is a descendant of a node $N =
<not(S_i),{}> \mif L_1,...,L_n$ in which, for each rule $r_j$ with
head $S_i$ in $P \cup I$, $L_j$ is a default conjugate of some literal
in the body of $r_j$.  (The situation is analogous to that depicted in
Figure~\ref{fig:schema}).  Now we consider the classes of goal
literals in the node $N$.  Those that are in $N$ but not in the
$DelayList$, were not resolved away via an {\sc Abduction} operation,
and were not resolved away via an \anscr{} resolution using an answer
with a non-abductive context form witnesses of unusability for some
rule for $S_i$ in $P \cup I$ by the same argument as in case (1a) in
the proof of Lemma~\ref{lem:co-unfounded-wfs}.  Similarly, those
literals in the $DelayList$ of $N_{leaf}$ form witnesses of
unusability by the same argument as in case (1b) of that proof.  This
leaves literals that were resolved away via {\sc Abduction}
operations, or were resolved away via an \anscr{} resolution with an
answer with a non-abductive context.  Either operation will union the
abductive context of a parent node with new objective literals to
produce a new child node.  We first note that by
Definition~\ref{def:co-unfounded}, the union of the contexts of all
answers in any co-unfounded set is consistent so that the single
abductive context, $C_i$, must also be consistent.  Let $O_i \in C_i$
be a given abducible objective literal.  In the first case, $O_i$ is
the explicit conjugate of some literal in a rule $r_j$ of $S_i$ in $P
\cup I$ added to $C_i$ directly through an {\sc Abduction} operation.
In the second case, $O_i$ is necessary to derive an answer that was
used via \anscr{} with a coherency axiom or other program or integrity
rules to remove a literal from $r_j$, and so form a witness of
unusability.  Since abductive contexts must be consistent, $O_i$ is
true in $I_{Context}$ (i.e. the interpretation induced by the union of
abductive contexts of all answers in $S^A$), iff it is a witness of
unusability for $r_i$ with respect to $\cI_{induced} \cup
I_{context}$.

\item Again, for the case of finite abductive scenarios, the argument
is essentially similar to that of Lemma~\ref{lem:co-unfounded-wfs},
but with the addition of abducibles.  The only difference is to ensure
that the union of contexts of all nodes in the co-unfounded set of
answers corresponding to $S^O$ is consistent, which fact follows from
Definition~\ref{def:ab-scen} which implies that the interpretation of
abducibles in an abductive scenario gives rise to a consistent
interpretation of these abducibles once their truth values are
propagated to default literals through coherency.

%---------------------------------------------------------------
\end{enumerate}
Extending the proofs to the case of infinite programs means taking
account of folding rules created by the use of the dual transformation
of Definition~\ref{dual-fold}, as discussed in
Lemma~\ref{lem:co-unfounded-wfs}.  It is straightforward to see that
the folding rules ensure do not affect consistency of abductive
contexts.  In addition, since abducibles are propagated through
folding rules (and all other rules), and since there can only be a
finite number of abducibles to be propagated, extension of the rest of
the argument is straightforward.
\end{proof}

%-------------------------------------------------------
We next prove a restricted form of Theorem \ref{thm:exdual}, which
assumes that the final forests exist for an \wfsmeth{} evaluation of a
query to these forests.  It uses the notion of the model $M(\sigma)$
of abductive solution $\sigma$ as introduced in
Definition~\ref{def:abd-sol}, and of a rule $query \mif{}
not(\bottom),Q$ as introduced in Definition~\ref{def:wfsmeth-eval}.
%-------------------------------------------------------
\noindent

\begin{lemma}	\label{lem:abdual-wfs}
Let $\cE$ be an \wfsmeth{} evaluation of a query $Q$ against an
abductive framework $< P,\cA,Int >$, whose final forest is
$\cE_\beta$.  Then $<query,Set>\mif | $ is an answer in $\cE_\beta$
iff $\sigma = < P,\cA,Set',I >$ is an abductive solution
such that $M(\sigma) \models Q$ and $Set \subseteq Set'$.
\end{lemma}

\noindent
\begin{proof} (Sketch) 
Given Lemmas~\ref{lem:co-unfounded-wfsab} and Lemma \ref{dual-wfs} the
proof is straightforward.  Soundness ($\Rightarrow$) is shown by an
induction on the length of the \wfsmeth{} evaluation, such that each
literal in the induced interpretation of the \wfsmeth{} forest is also
in the well-founded model.  Completeness ($\Leftarrow$) is shown by a
double induction on construction of $M(\sigma)$, the well-founded
model of an abductive scenario with the the operators of
Definition~\ref{def:dualintops} used in the inner induction, and the
operator of Definition~\ref{def:dualfp_inner} used in the outer
induction.
\end{proof}

%------------------------------------------------------

\noindent
{\bf Theorem \ref{thm:exdual}} 
{\em 
Let $< P, \cA, I >$ be an abductive framework, and $\cE$
be an \wfsmeth{} evaluation of $Q$ against $< P, \cA, I
>$.  Then
\begin{enumerate}
\item $\cE$ will have a final forest $\cE_{\beta}$;
\item if $<query,Set>\mif $ is an answer in $\cE_\beta$ $\sigma
= < P,\cA,Set,I >$ is an abductive solution for $< P,
\cA, I >$;
\item if $< P,\cA,Set,I >$ is a minimal abductive solution
for $Q$, then $<query,Set>\mif $ is an answer in $\cE_\beta$.
\end{enumerate}
}
\noindent
\begin{proof}
(Sketch) 
\begin{enumerate}
\item The first statement follows from an argument similar to that
made for extended SLG trees in \cite{Swif99b}.  Briefly recapitulated,
it can be seen that all \wfsmeth{} trees are of finite depth,
therefore they must have at most a countably infinite number of nodes
(e.g. see H. Rogers, Theory of Recursive Functions and Effective
Computations, MIT press, 1987 Section 16.3).  Furthermore there are at
most a countably infinite number of \wfsmeth{} trees in a \wfsmeth{}
forest.  Now at each successor ordinal of a \wfsmeth{} evaluation,
each \wfsmeth{} operation either creates a new tree or adds at least
one node to an existing tree.  At each limit ordinal the union of all
forests indexed by lower ordinals is taken.  Therefore, a \wfsmeth{}
evaluation can have at most a countably infinite number of states.
Thus, the ordinal $\beta$ is reachable via the transfinite induction
of Definition \ref{def:wfsmeth-eval}.
\item Both the second and third statements are immediate from
Lemma~\ref{lem:abdual-wfs}. 
\end{enumerate}
\end{proof}

\subsection{Proof of Theorems in Section \ref{sec:term-compl}} 

%-----------------------------------------------------------------
\mycomment{
\begin{definition} [Bounded-term-size
Property] The size of a term is defined recursively as follows:
\begin{itemize} 
\item The size of a variable or constant is 1 
\item The size of a compound term $f(t_1,...,t_n)$ is 1 plus the sum
of the sizes of its arguments. 
\end{itemize}

A finite program $P$ has the bounded-term-size property if there is a
function $f(n)$ such that for any forest $\cF$ of any \wfsmeth{}
evaluation $\cE$, whenever an atomic query $Q$ has no
arguments whose sizes exceed $n$, no atom in any node of $\cF$ has
size greater than $f(n)$.
\end{definition}
}
\mycomment{
\begin{definition}
Let $P$ be a finite program, and $P_I$ its intension (Definition
\ref{def:data-complexity}).  Then $|P|$ denotes the number of rules in
$P$, and $\Pi_P$ denotes the maximum number of literals in the body of
a rule in $P$.  Let $s$ be an arbitrary positive integer.  Then
$\cN(s)$ denotes the number of atoms of predicates in $P$ that are not
variants of each other and whose arguments do not exceed $s$ in size.
\end{definition}
}
%-----------------------------------------------------------------

\begin{definition} \label{def:size}
Let $P$ be a program or dual program that is finite and ground.
$size(P)$ denotes the sum, for each rule $R$ in $P$ of 1 plus the
number of body literals in $R$; $size(P|_L)$ denotes the size of rules
in $P$ whose head is the literal $L$.  $heads(P)$ denotes the set of
literals that occur as heads of rules in $P$.
\end{definition}

%-----------------------------------------------------------------------
\mycomment{
Then $|P|$ denotes the number of rules
in $P$, $\Pi_P$ denotes the maximum number of literals in the body of
a rule in $P$, 
}

\mycomment{
\begin{definition} \cite{VRS91} \label{def:data-complexity}
The {\em intension} of a program $P$, $P_I$, consists of all
rules in $P$ with non-empty bodies; the {\em extension} of $P$,
$P_E$, consists of all rules in $P$ whose body is empty.  The data
complexity of $P$ is the computational complexity of deciding an
answer to a ground atomic query as a function of the size of $ P_E$.
\end{definition}
}
%-----------------------------------------------------------------------

The following lemma shows that the size of the program produced by the
dual transformation is linear in the size of the original program.
The bound provided is not the tightest possible.

\begin{lemma} \label{dual-sizes}
Let $P$ be a finite ground extended program and $\cA$ a finite set of
abducibles.  Then $size(dual(P \cup \cA)) < 9 size(P) + 2|\cA|$.
\end{lemma}
\begin{proof}
Let $O$ be an objective literal for which there are $m > 0$ rules in
$P$ with total size $size(P|_O)$.  
\begin{enumerate}
\item  Case (1) of Definition~\ref{dual-fold}.  There will be $m + 2$
rules with $fold^a_i\_O$ in their heads or bodies produced by case
(1a) of Definition~\ref{dual-fold} for a total size of $3m +4$.  The
total size of rules with heads of the form $not(fold^b_i\_O)$ produced
by (1b) will be $2(size(P|_O) - m)$, so that the total number of
folding rules for $O$ will be $2size(P|_O) + m +4$.  Summed up over
rules for all objective literals in $P$, this is $2size(P) + 4
heads(P) + rules(P)$, where $rules(P)$ is the number of rules in $P$.

\item Case (2).  The size of the rules produced by case (2) is bounded by
$2(literals(P) - heads(P))$.  

\item Case (3) Finally, the size of the axioms of coherence is bounded
by $|literals(P)| + 2|\cA|$, where $|literals(P)|$ is the number of
literals in $P$.
\end{enumerate}

Note that $|literals(P)| \leq size(P)$, as is $|heads(P)|$ and
$|rules(P)|$, so that $size(dual(P)) < 9 size(P) + 2|A|$. 
\end{proof}

\begin{lemma} \label{tot-ops}
Let $P$ be a finite ground extended program, and $< P,\emptyset,
\emptyset >$ be an abductive framework.  Let $\cE$ be an
\wfsmeth{} evaluation of a non-abductive query $Q$ against $dual(P)$,
whose final forest is $\cF_{\beta}$.  Then $\cF_{\beta}$ can be
constructed in at most $2size(dual(P)) + literals(P)$ steps.
\end{lemma}
\begin{proof}
It takes at most one \wfsmeth{} operation to create a node in
$\cF_{\beta}$: thus the number of nodes in this forest is an upper
bound on the number of \wfsmeth{} operations required to evaluate $Q$.
In $\cF_{\beta}$ there is at most one tree for each literal in
$literals(dual(P))$.  The root node $N_L$ for a literal $L$ in
$dual(P,\emptyset)$ has one child for each rule for $L$ in $P$.
Consider a child $N_R$ of $N_L$ formed by \pgmcr{} using a rule $R$.
Then there are at most $2L_R$ descendants of $N_R$ in $\cF_{\beta}$
where $L_R$ are the number of body literals in $R$.  To see this,
first note that the number of goal literals in $N_R$ is $L_R$.
Further, since $dual(P,\emptyset)$ is ground and the set of abducibles
is empty, any descendant $N_D$ of $N_R$ can have at most one child for
each of the operations in Definition \ref{def:ops}.  Also consider
that any child $N_{child}$ of $N_D$
\begin{enumerate}
\item has the same number of goal literals as $N_D$ and fewer
delay literals; (e.g. if $N_{child}$ was produced by \simpl{} or \unfr
); or
\item has the same number of delay literals as $N_D$ and one fewer
goal literal (e.g. if $N_{child}$ was produced by \anscr ); or 
\item has one more delay literal than $N_D$ and one fewer goal
literal than $N_D$ (e.g. if $N_{child}$ was produced by \delay ); or
\item is a failure node (e.g. if $N_{child}$ was produced by \simpl ).
\end{enumerate}
In cases (1) and (2) above the size of $N_D$ is reduced; in case 3 we
note that by Definition \ref{def:ops} a literal can be delayed only
once in any path from a root of a tree to a leaf.  In case 4, we note
that no operations are applicable to a failure node.  Thus the length
of the path from $N_O$ to the unique leaf descendant of $N_R$ in
$\cF_{\beta}$ is at most $2L_R$, leading to the bound in the
statement.
\end{proof}

%-------------------------------------------------------------------
\noindent
{\bf Theorem \ref{termination} } {\em Let $< P,\cA,I
>$ be an abductive framework such that $P$ and $I$ are finite
ground extended programs, and $\cA$ is a finite set of abducibles.  Let
$\cE$ be an \wfsmeth{} evaluation of a query $Q$ against $< P,\cA,I
>$.  Then $\cE$ will have a final forest that is produced by a
finite number of \wfsmeth{} operations.}

\noindent
\begin{proof}
By Lemma~\ref{dual-sizes}, $dual((P \cup I),\cA)$ is finite if
$< P, \cA, I >$ is finite, so we need only consider the
direct evaluation of $dual((P \cup I),\cA)$.  In the case in which
$\cA$ is empty, the statement is immediate from Lemma
\ref{tot-ops}.  In the case in which $|\cA| = n$, $n > 0$, then the
{\sc abduction} operation must be taken into consideration.  As in the
proof of Lemma~\ref{tot-ops}, the total number of trees in any forest
of $\cE$ is bounded by $literals(dual((P \cup I), \cA))$.  Let $N_L$
be the root node of a tree for $L$ in a forest $\cF$ of $\cE$, and
$N_R$ be a child of $N_L$ produced by \pgmcr{} using a rule $R$.  As
in the proof of Lemma~\ref{tot-ops}, path to any descendant of $N_R$
is at most 2$L_R$, where $L_R$ is the number of body literals in $R$.
This is immediate Lemma~\ref{tot-ops} since the {\sc abduction}
operation reduces the number of goal literals in any node by 1.
However, unlike the case in Lemma~\ref{tot-ops} where the set of
abducibles is empty, the number of children of $L_R$ or its
descendants can be greater than one, depending on the number of
abductive solutions for a goal literal or set of delay
literals. However, the number of these solutions is always finite, so
that $\cF$ will have a finite number of nodes.  As before, each node
is produced by a single \wfsmeth{} operation, so that $\cF$ can be
produced by a finite number of operations.  Since an arbitrary forest
$\cF$ must be finite, a final forest must be produced by a finite
number of \wfsmeth{} operations.
\end{proof}

%-------------------------------------------------------------------

\begin{theorem}
Let $\cF$ be the final forest in a \wfsmeth{} evaluation $\cE$ of a
query $Q$ against an abductive framework $< P, \cA, I >$.
Let $C_{context}$ be the maximal cardinality of the context of any
abductive subgoal in $\cF$, and $C_{abducibles}$ be the cardinality of
$\cA$.  Then $\cF$ can be constructed in at most $M \times
2size(dual((P \cup I),\cA )$ steps, where 
\[
M = \sum_{i \leq C_{context}} \left( \begin{array}{c}
					C_{abducibles} \\ i
				     \end{array}
			      \right)
\]
\end{theorem}
\begin{proof}
Let $N_L$ be a root node of a tree for a literal $L$ in $\cF$.  As
noted in the proof for Lemma~\ref{tot-ops}, there is one child for
$N_L$ for every rule $R$ for $L$ in $dual((P \cup I), \cA)$.  Such a
node, $N_R$ has a number of leaf descendants that is bounded by the
number of abductive solutions that are possible to return, via \anscr
, to the descendants of $N_R$.  By construction, this bound is $M$.
Again by the considerations in Lemma~\ref{tot-ops}, the length of the
path from $N_R$ to any leaf is at most $2R_L$ where $R_L$ is the
number of body literals in the rule that produced $N_R$.  Further, the
length of the path from $N_L$ to any leaf is $2R_L + 1$, which is 2
times the size of $R$ minus 1. so that the number of nodes in the tree
rooted by $N_L$ is bounded by $2 M size(P|_L)$.  Summing this for all
literals in $dual((P \cup I),\cA)$, the bound of the statement is
obtained.
\end{proof}

\bibliographystyle{acmtrans}
\bibliography{longstring,all,newrefs}
%\newpage

\end{document}